\documentclass{article}
\usepackage[utf8]{inputenc}
\usepackage{lipsum}
\usepackage[T1]{fontenc}
\usepackage[english]{babel}
\usepackage{geometry}
\usepackage{tikz}
\usepackage{booktabs}
\usepackage{float}
\usepackage[english]{varioref}
\usepackage{amsmath,amsfonts,amsthm,amscd,amssymb}
\usepackage{graphicx}
\usepackage{sidecap}
\usepackage[font=small,hang]{caption}
\usepackage{subfig}
\usepackage{url}
\usepackage{multicol}
\usepackage{hyperref}
\usepackage{authblk}

\title{Maximum entropy approaches for the study of triadic motifs in the Mergers \& Acquisitions network}
\author[1,2]{Ihusan Adam}
\author[3]{Stefano Garlaschi}
\author[4,5]{Jian-Hong Lin}
\author[6,7]{Simone Piaggesi}
\author[8]{Matteo Barigozzi}
\author[9]{Andrea Gabrielli}
\author[10]{Rossana Mastrandrea}

\affil[1]{\footnotesize Department of Information Engineering, University of Florence, Italy}
\affil[2]{\footnotesize Department of Physics and Astronomy, University of Florence, Italy}
\affil[3]{\footnotesize Department of Physics and Astronomy "Galileo Galilei", University of Padova, Italy}
\affil[4]{\footnotesize URPP Social Networks, University of Zurich, Andreasstrasse 15, CH-8050 Z\"urich, Switzerland}
\affil[5]{\footnotesize ETH Z\"urich, Department of Management, Technology and Economics, Z\"urich, Switzerland}
\affil[6]{\footnotesize Department of Computer Science and Engineering, University of Bologna, Italy}
\affil[7]{\footnotesize ISI Foundation, Torino, Italy}
\affil[8]{\footnotesize Department of Statistics, London School of
Economics and Political Science, London, England.}
\affil[9]{\footnotesize Istituto dei Sistemi Complessi (CNR) UoS Sapienza, Dipartimento di Fisica, Sapienza Università di Roma, Italy}
\affil[10]{\footnotesize IMT School for Advanced Studies, Lucca, Italy}

\date{}

\begin{document}

\maketitle
\begin{abstract}
In the past years statistical physics has been successfully applied for complex networks modelling. In particular, it has been shown that the maximum entropy principle can be exploited in order to construct graph ensembles for real-world networks which maximize the randomness of the graph structure keeping fixed some topological constraint. Such ensembles can be used as null models to detect statistically significant structural patterns and to reconstruct the network structure in cases of incomplete information.

Recently, these randomizing methods have been used for the study of self-organizing systems in economics and finance, such as interbank and world trade networks, in order to detect topological changes and, possibly, early-warning signals for the economical crisis. In this work we consider the configuration models with different constraints for the network of mergers and acquisitions (M \& As), 
Comparing triadic and dyadic motifs, for both the binary and weighted M\&A network,  with the randomized counterparts can shed light on its  organization at higher order level. 


\end{abstract}

\section{Introduction}

In recent years the study of complex networks has strongly interconnected an increasing interest \cite{cimini_statistical_2019,milo_network_2002}. Our modern global society can be naturally modeled within this framework \cite{Albert2001StatisticalMO,Newman2003TheSA,article,bargigli_random_2011}. The advantage of this approach allows to study real-world phenomena with simple tools coming from Statistical Physics field and able to reveal fundamental intrinsic properties about the topological organization and its macroscopic effects. \cite{hutchison_triadic_2012,duenas_spatio-temporal_2017,squartini_early-warning_2013}.

One topic that has received a lot of attention concerns the possibility to use null models to reconstruct network  from partial information or test the validity of its binary and weighted local and global properties. A lot of efforts have been addressed in the research of a procedure to reconstruct the network structure starting from the available information\cite{milo_network_2002,squartini_analytical_2011,squartini_unbiased_2015}. In this way it is possible to obtain null models that can be tested with respect to the data.

Economic and financial markets are among the most investigated systems thanks to their impact on our globalized society \cite{hutchison_triadic_2012,duenas_spatio-temporal_2017,in_t_veld_finding_2014,bargigli_random_2011}. The mathematical and computational framework of network science provides a suitable set of quantitative tools with which we can naturally model them and probe their status.

As shown in \cite{squartini_early-warning_2013}, starting from data referring to the time window $1998$–$2008$ for interbank exposures of Dutch banks, comparing  the predictions obtained from the null models with the empirical evidences, it was possible to detect anomalies in the topology of higher-order patterns in correspondence to economic crisis periods.

Inspire by that, in this work we study the triadic motifs of the M\&A network which represents an interesting example of the global interactions among world economies. 

\section{Data}

In this paper, we use data extracted  from  the Worldwide Mergers, Acquisitions, and Alliances Databases SDC Platinum (Thomson Reuters). The financial databases provide extensive large-scale information on global transactions since 1985 to 2010. Most of the recorded transactions refer to domestic M\& As-activity (74\%, on average). The domestic links represent around 10\% which indicats that, on average, M\&As tend to be more oriented towards abroad, but the average volume of foreign operations is much lower. We focus only on the period 1995-2010 because earlier years are highly characterized by missing information. At each year, the international merger and acquisition network could be described as a directed weighted network $G=(V,E)$, with $n=|V|$ nodes and $m=|E|$ links, where $V$ is the set of the nodes and $E$ is the set of links. The weighted network is described as a weighted matrix $\bold{W}$, where $w_{ij}$ represents the amount of amount of M\&A from node $i$ and to $j$. We define the directed unweighted network in terms of the binary adjacency matrix $\bold{A}$, where $a_{ij}=1$ if $w_{ij}>0$, otherwise $a_{ij}=0$.  


In the following analysis we neglect the transitions between companies belonging to the same nation. This means that we ignore self-connections and so we can arbitrarily fix the diagonal of the matrix $\bold{A}$, setting all its elements equal to $0$ for later convenience. Furthermore, we have to be sure to work with a connected network, this means there cannot be isolated sub-components. For this reason, for each yearly snapshot, we take into account only the data referring to the giant component.


\section{Methods}
The purpose of the paper concerns finding motifs, subgraphs with size 3 in the world web of Mergers and Acquisitions (M\&A) dataset comparing year by year the real-world network with null models which preserve some topological properties obtained from the data while maximizing the randomness of all the other features.

In the first place we look just at the binary structure of the networks, later on we consider also it intrinsic weighted nature.


 


\subsection{Maximum Entropy approach}

\begin{figure}
    \centering
    \includegraphics[width=0.5\textwidth]{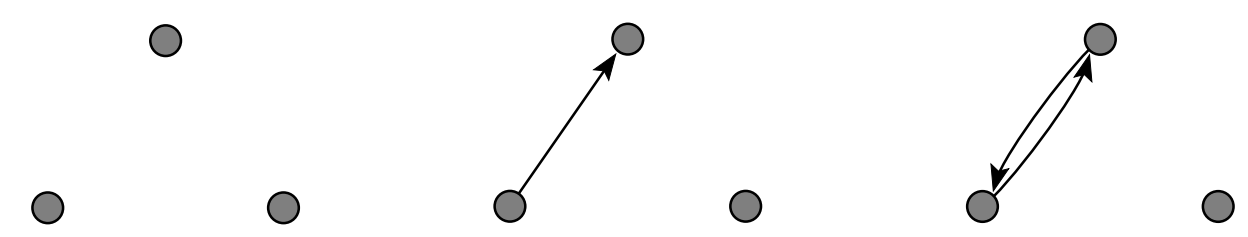}
    \caption{Caption}
    \label{fig:dyadic-motifs}
\end{figure}

\begin{figure}
    \centering
    \includegraphics[width=0.5\textwidth]{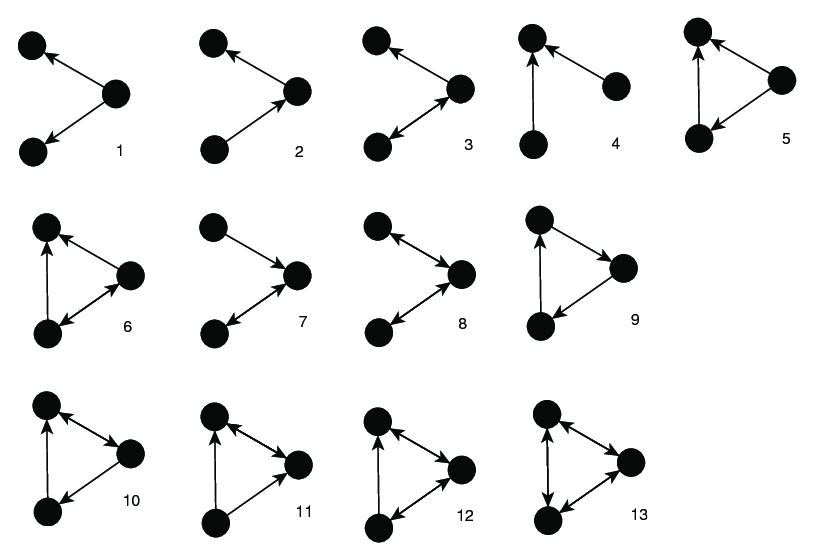}
    \caption{The 13 possible directed triadic motifs that can occur.}
    \label{fig:triadic-motifs}
\end{figure}

From the data we can count the number of times that a certain topological motif occurs for each year. By motif we mean directed subgraphs and we focus on the detection of the dyadic and triadic ones, i.e. the motifs made up by two or three nodes as shown in Fig. \ref{fig:dyadic-motifs} and  \ref{fig:triadic-motifs}. 

\begin{figure}
    \centering
    \includegraphics[width=0.5\textwidth]{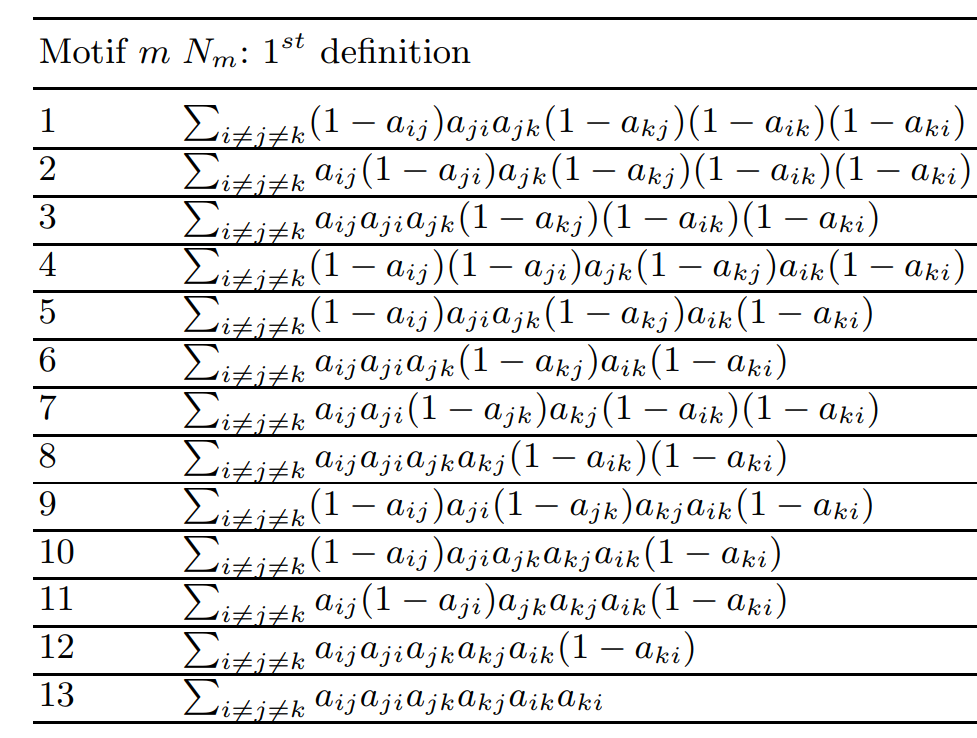}
    \caption{Here are displayed the formulas that have to be used, once the binary adjacency matrix is given, to count the number of times one of the 13 motif occurs.}
    \label{fig:tab}
\end{figure}

To do so in the case in which we just focus on the binary structure of the network we just count them as explained in Figure \ref{fig:tab}.

In the weighted network scenario instead, among all the possible choices of defining the motifs we choose the simplest one. We look again at the presence of the motifs in the binary sense, but now every time that a certain motif is detected we assign to it a weight that is simply the product of the weights of the links that compose the subgraph.

Null models used in this work belong to the class of the maximum entropy approaches. This method prescribes to search for  the probability distribution $P(A)$ that maximize the Shannon entropy \cite{jaynes_information_1957,bianconi_entropy_2007}:
\begin{equation}
S\left[P(A)\right]=- \sum_{A} P(A) \ln P(A)
\end{equation}
imposing the normalization constraint, $ \sum_{A} P(A)=1$. In these null models other additional constraints can be taken into account, leading to different null models. In the first approach we just fix the in-degree and out-degree . This leads to the so called \emph{Directed Configuration Model} (DCM). Another possibility is to  add also the conservation of the reciprocate degree, finding the \emph{Reciporcal Confiugration Model} (RCM). In the weighted case instead we want to constraint the in and out degrees and in and out strengths. 

Here, we are adopting a canonical approach, with respect to the constraints we are imposing do not need to hold for each configuration, but they have to be satisfied on average over the network ensemble.




From this constrainted maximization in the DCM we find that the probability of having a link from node $i$ to node $j$  is
\begin{equation}
p_{ij}=\frac{x_i y_j}{1+x_i y_j}
\end{equation}
where $x_i$ and $y_j$ are the hidden variables associated, respectively, to the out- and in-degree. 

In the RCM instead we constraint not only the node in aut degree, but also the total number of reciprocated links per node. Therefore, we need to distinguish the existence probability of link accordingly to the following fourth possibilities: 
\begin{equation}
\begin{cases}
p_{ij}^{\rightarrow}=\frac{x_i y_j}{1+x_i y_j +x_j y_i +z_i z_j} \\
p_{ij}^{\leftarrow}=\frac{x_j y_i}{1+x_i y_j +x_j y_i +z_i z_j} \\
p_{ij}^{\leftrightarrow}=\frac{z_i z_j}{1+x_i y_j +x_j y_i +z_i z_j}\\
p_{ij}^{\nleftrightarrow}=\frac{1}{1+x_i y_j +x_j y_i +z_i z_j}
\end{cases}
\end{equation}
where $x_i$, $y_j$ and $z_i$ are the hidden variables associated, respectively, to not-reciprocated node out and in-degree and the reciprocated node degree. 

Finally for the weighted case  the weights of each link in the synthetic networks are estimated by
\begin{equation}
\tilde{w}_{ij}=\frac{s_i^{out} s_j^{in}}{W p_{ij}}\tilde{a}_{ij}
\end{equation}
where $p_{ij}$ is the link probability according to the Fitness-induced Configuration Model (FiCM in \cite{Cimini2015SystemicRA}) , $s_i^{out}$ and $s_j^{in}$ are the observed out- and in-strengths, $W$ is the total observed weight and $a_{ij}$ stands for a particular realization of the reconstructed link.

\section{Results}
\subsection{Preliminary Analysis}
\begin{figure}[ht]
	\begin{center}
	\subfloat[]
	{\includegraphics[scale=0.3]{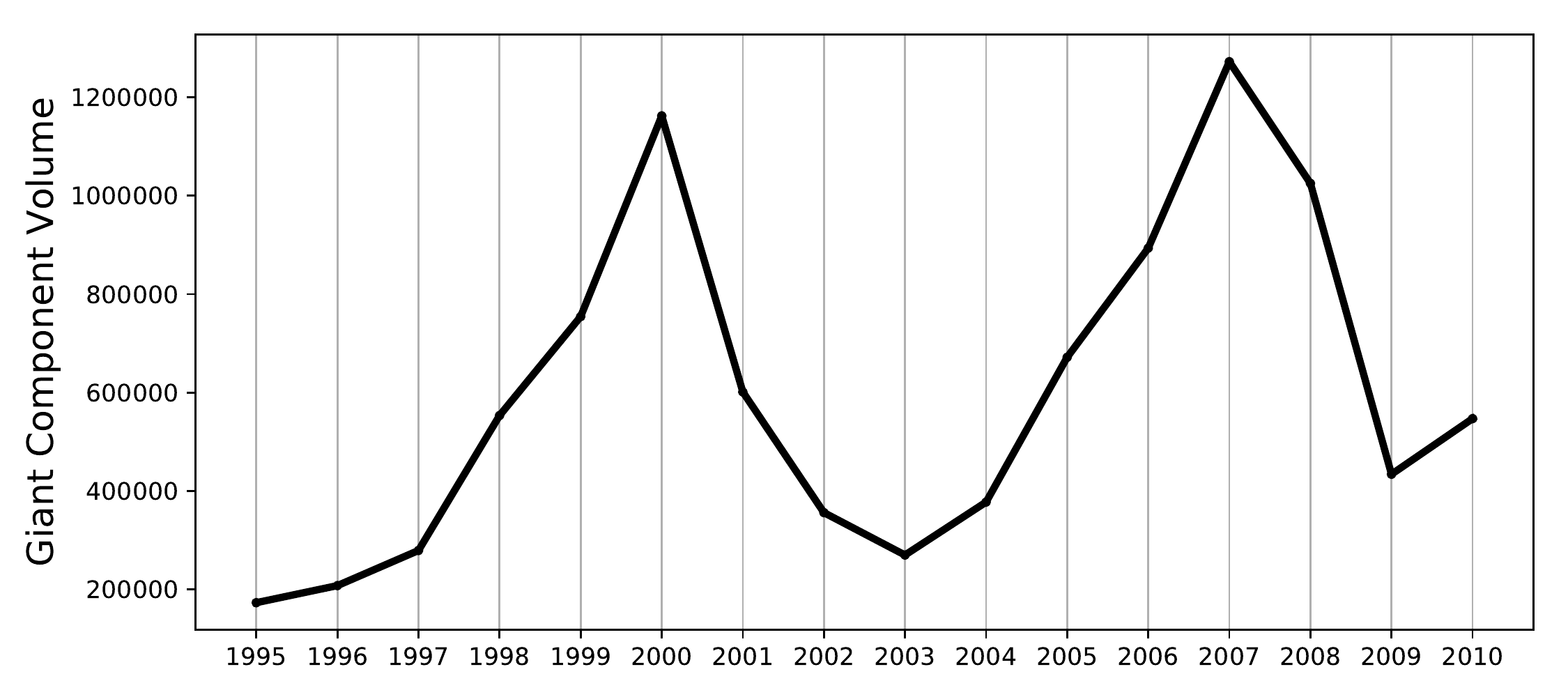}}
    \subfloat[]
	{\includegraphics[scale=0.3]{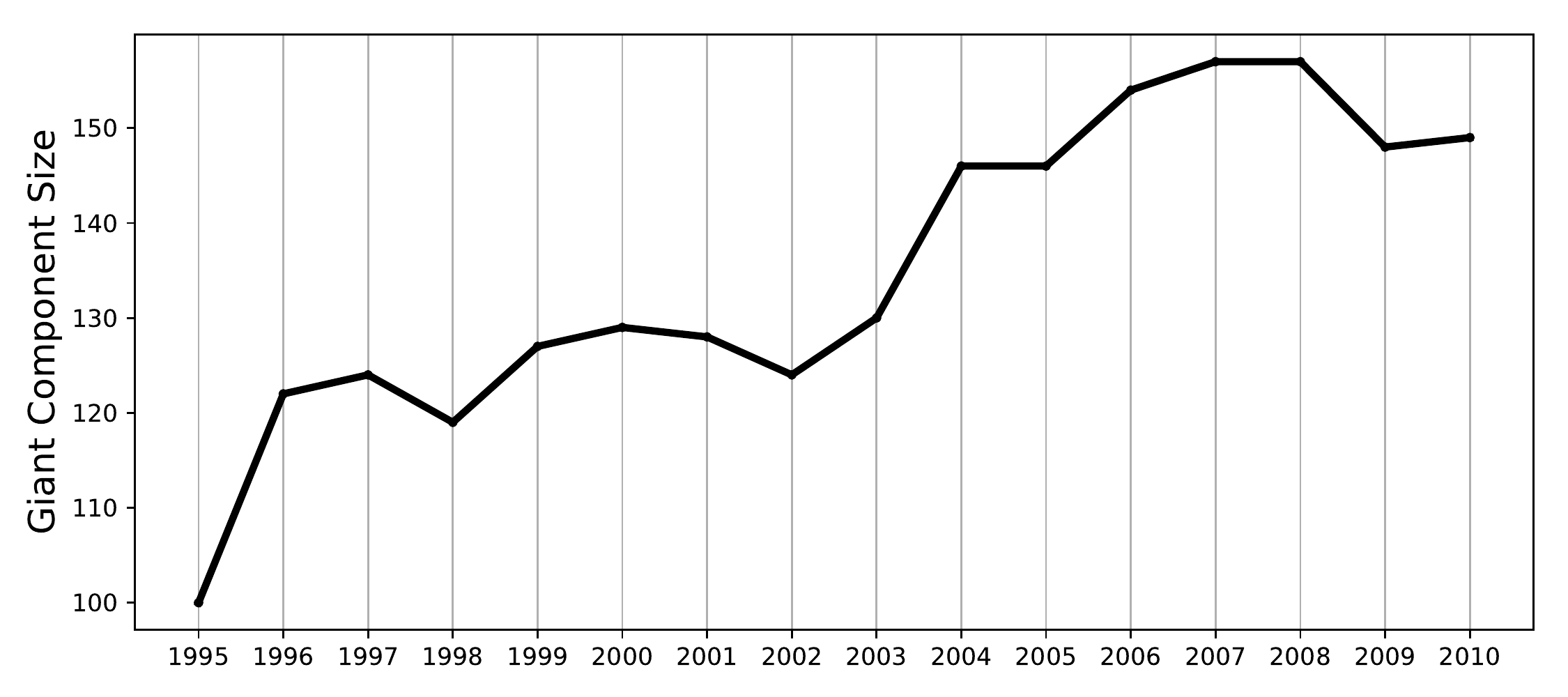}}\quad
    \subfloat[]
    {\includegraphics[scale=0.3]{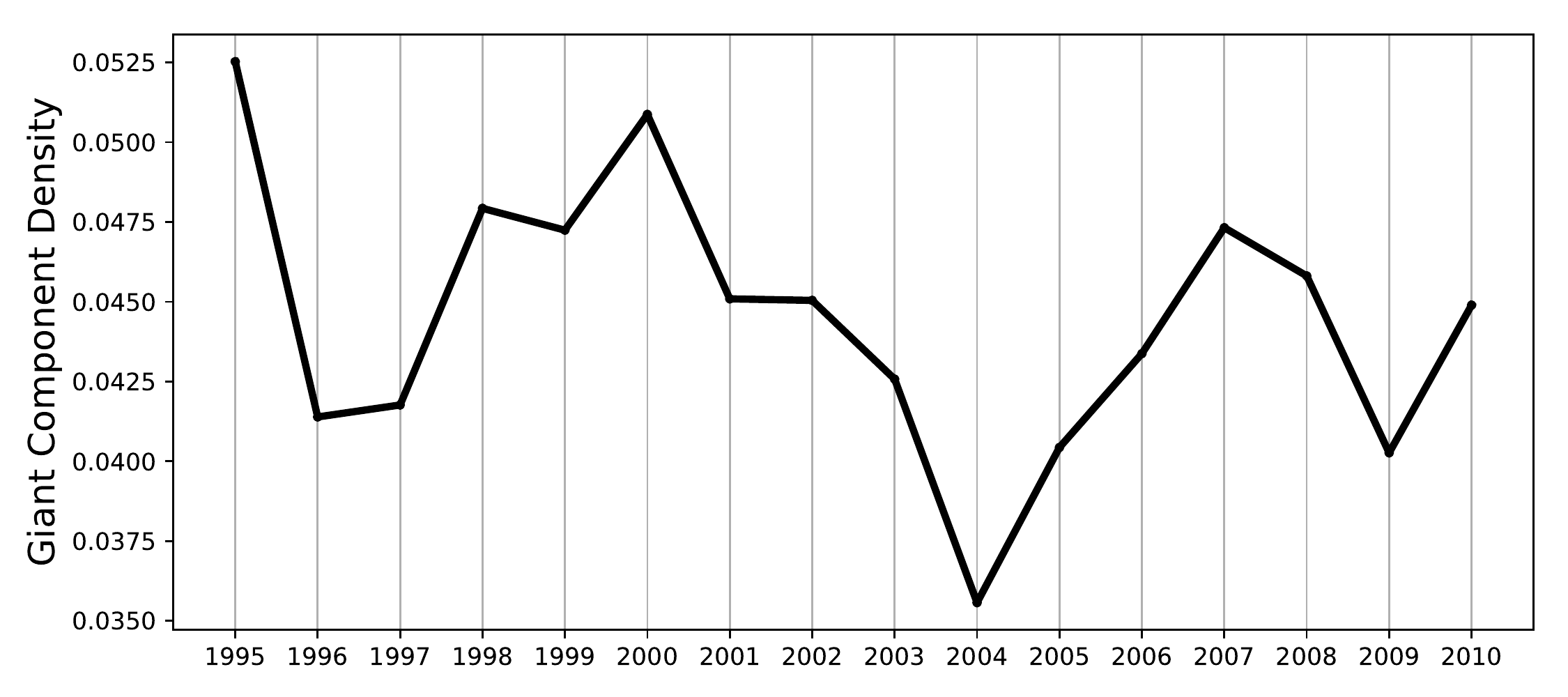}}
    \subfloat[]
	{\includegraphics[scale=0.3]{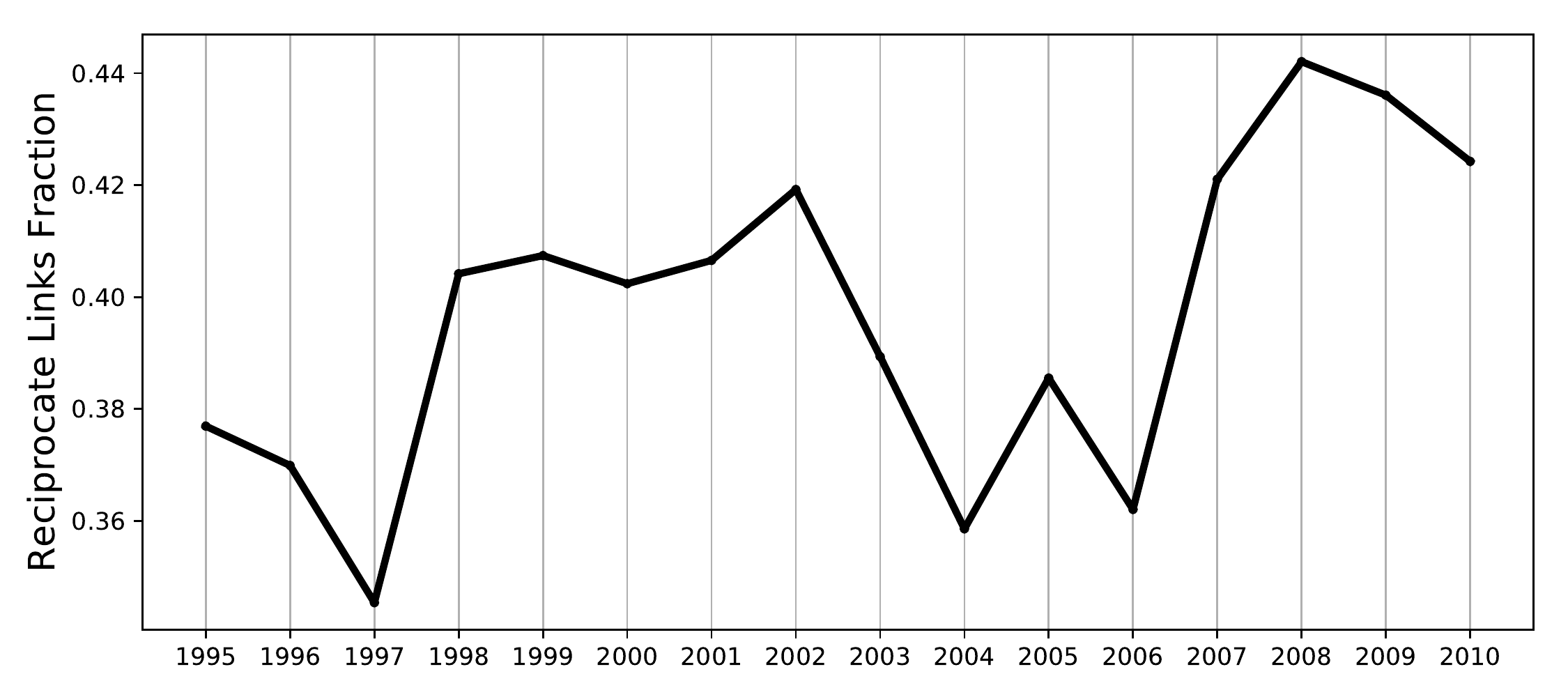}}
    \end{center}
    \caption{The giant component volume (a), size (b), density (c), and the reciprocate links fraction (d) of the real network at each yearly snapshot.}
	\label{fig:size}
\end{figure}

As a preliminary analysis, we analyzed the basic topology of the real network at each yearly snapshot. Figure~\ref{fig:size}(a) shows the giant component volume at each year, where the volume ranges from 200000 to 1200000. As shown in Fig.~\ref{fig:size}(b), the giant component size increases from 100 to 157. Whereas the giant component density, shown in Fig.~\ref{fig:size}(c) is below 0.0525, which suggests that for each year, the giant component is a sparse network.
Figure~\ref{fig:size}(d) shows the reciprocate links fraction of the real network at each snapshot, where the fraction ranges from 0.34 to 0.44.

\captionsetup[subfigure]{labelformat=empty}
\subsection{Motifs Pattern Analysis}
In this section we are going to analyze the statistical significance of co-occurrencies of different triadic motifs inside each time window of the empirical network data. In order to perform this task we compute z-score statistics of the observed number of a particular motif respect to the average value of counts coming from the null ensembles (both binary and weighted ones). We put a threshold of 3 standard devations as significance level. 

\begin{figure}[h]
	\begin{center}
    \subfloat[Directed Configuration Model]
	{\includegraphics[scale=0.3]{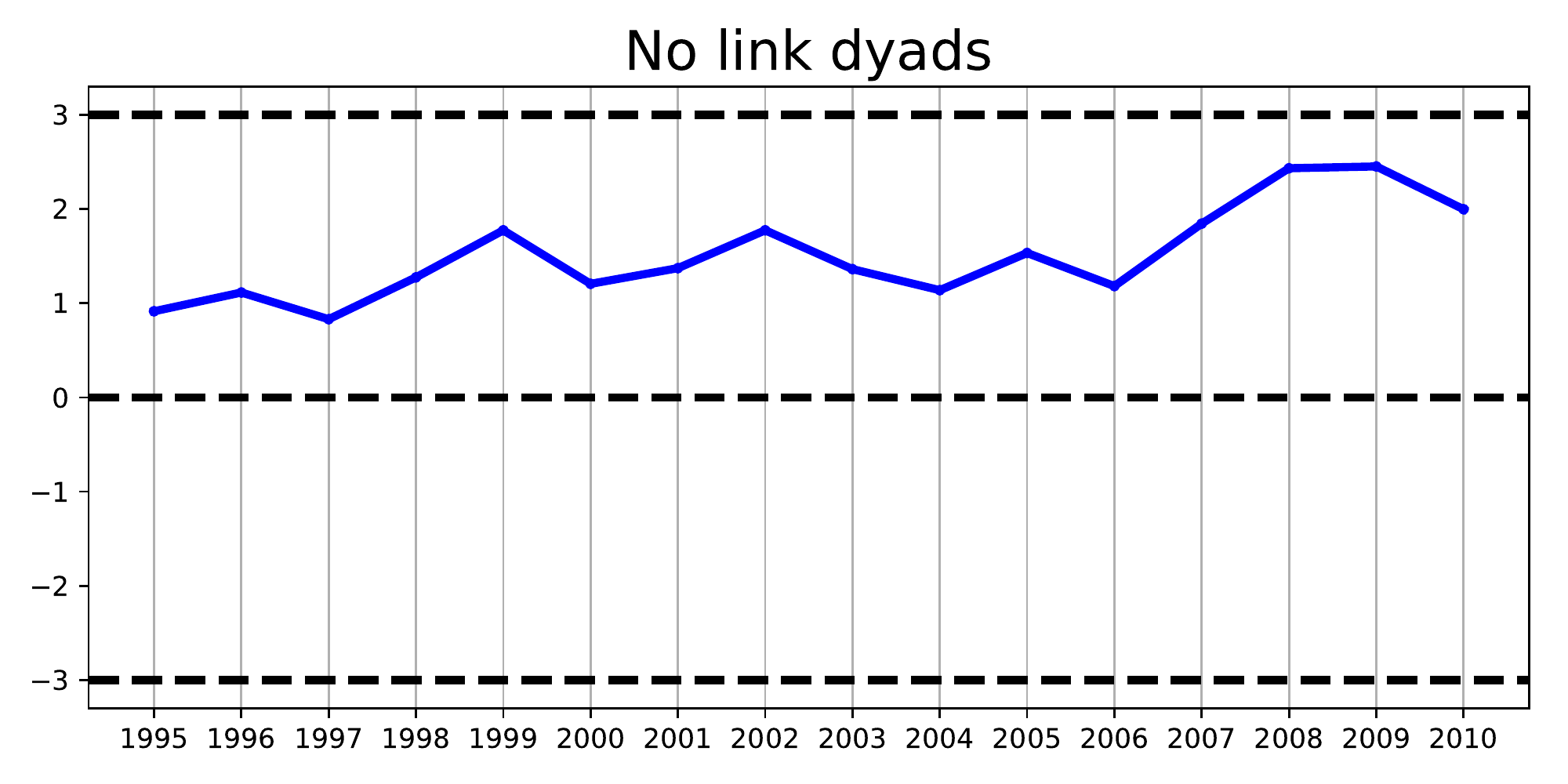}}
    \subfloat[Reciprocal Configuration Model]
    {\includegraphics[scale=0.3]{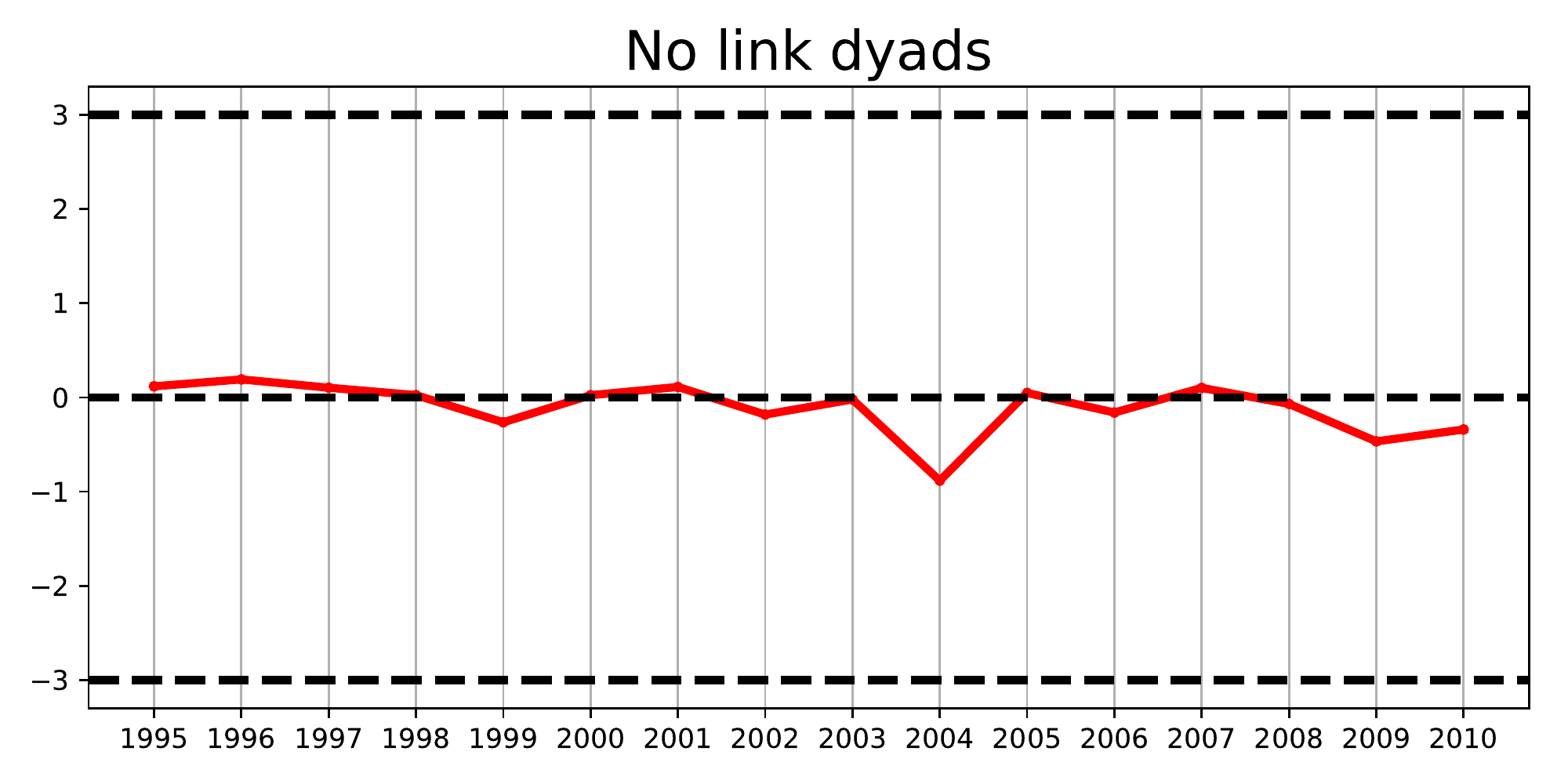}}\quad
    \subfloat[]
	{\includegraphics[scale=0.3]{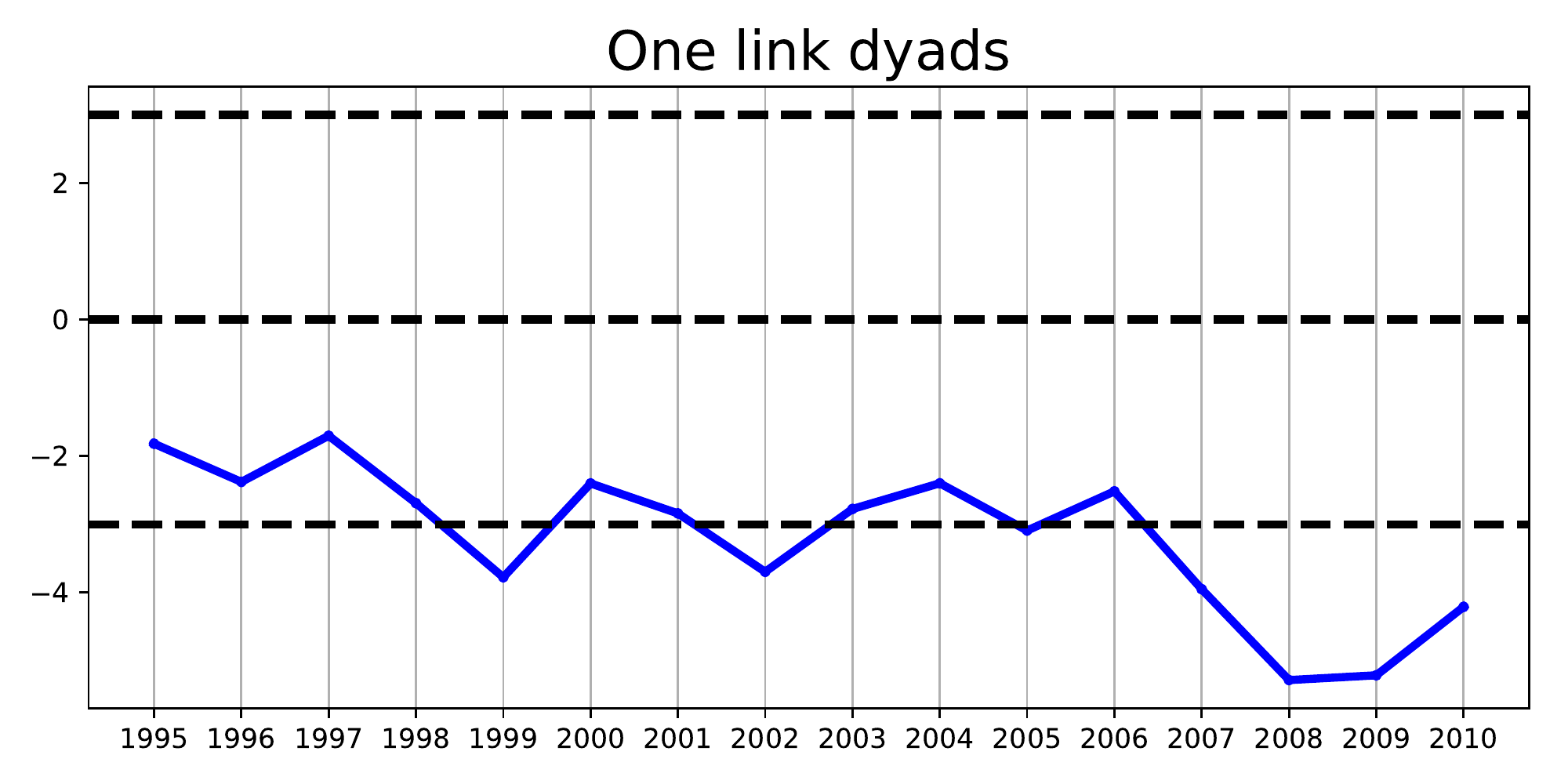}}
    \subfloat[]
    {\includegraphics[scale=0.3]{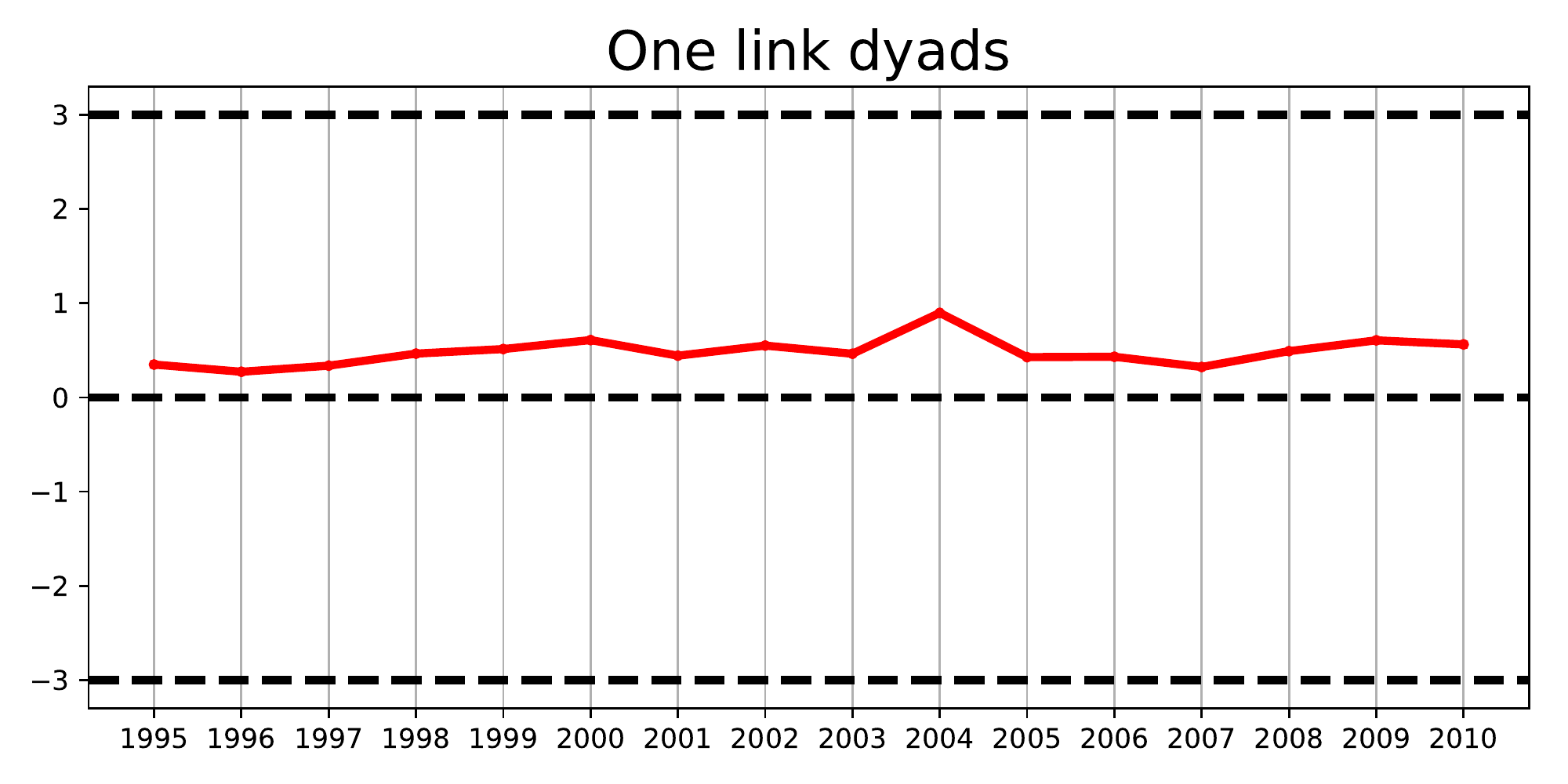}}\quad
    \subfloat[]
	{\includegraphics[scale=0.3]{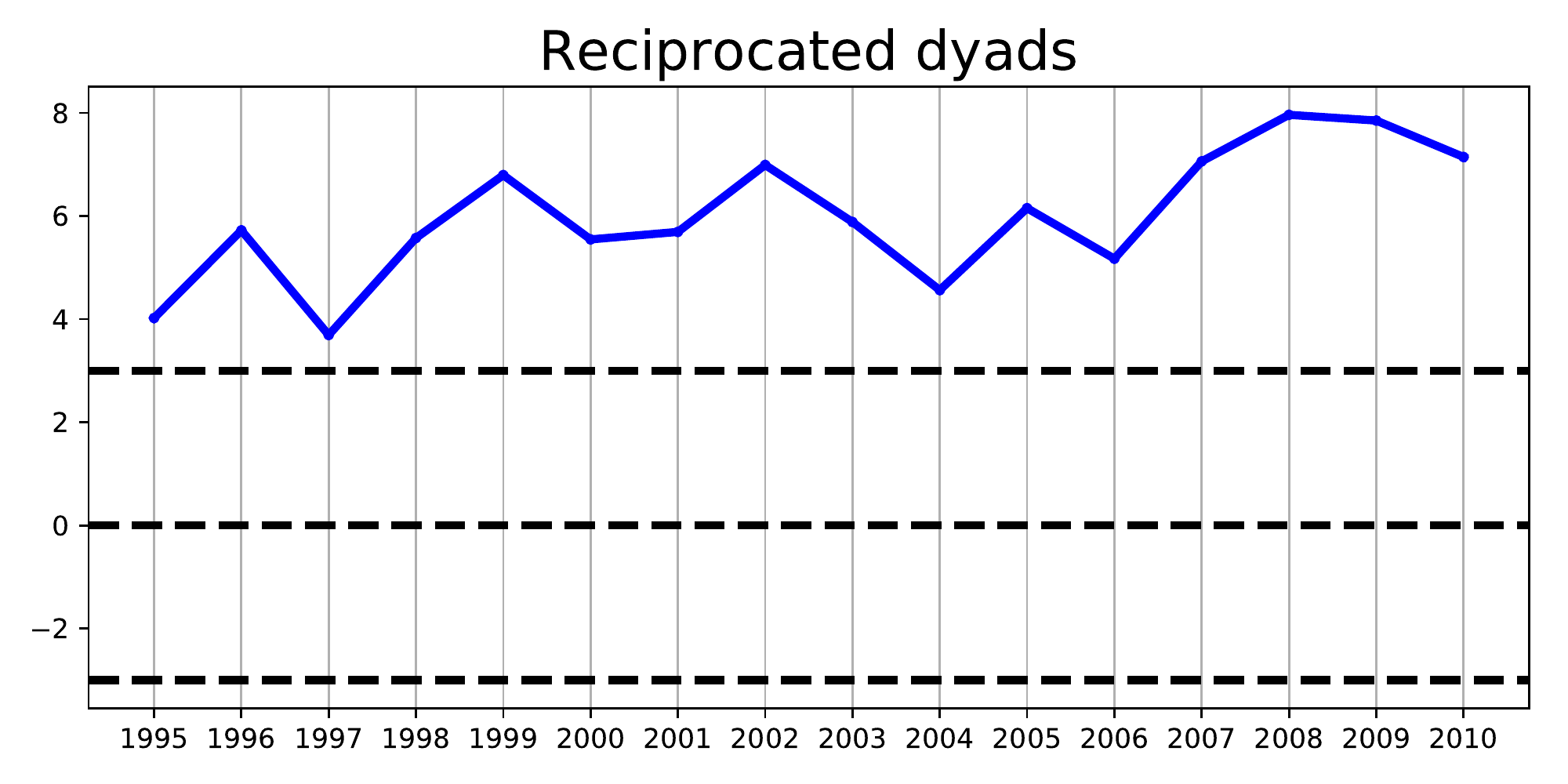}}
    \subfloat[]
    {\includegraphics[scale=0.3]{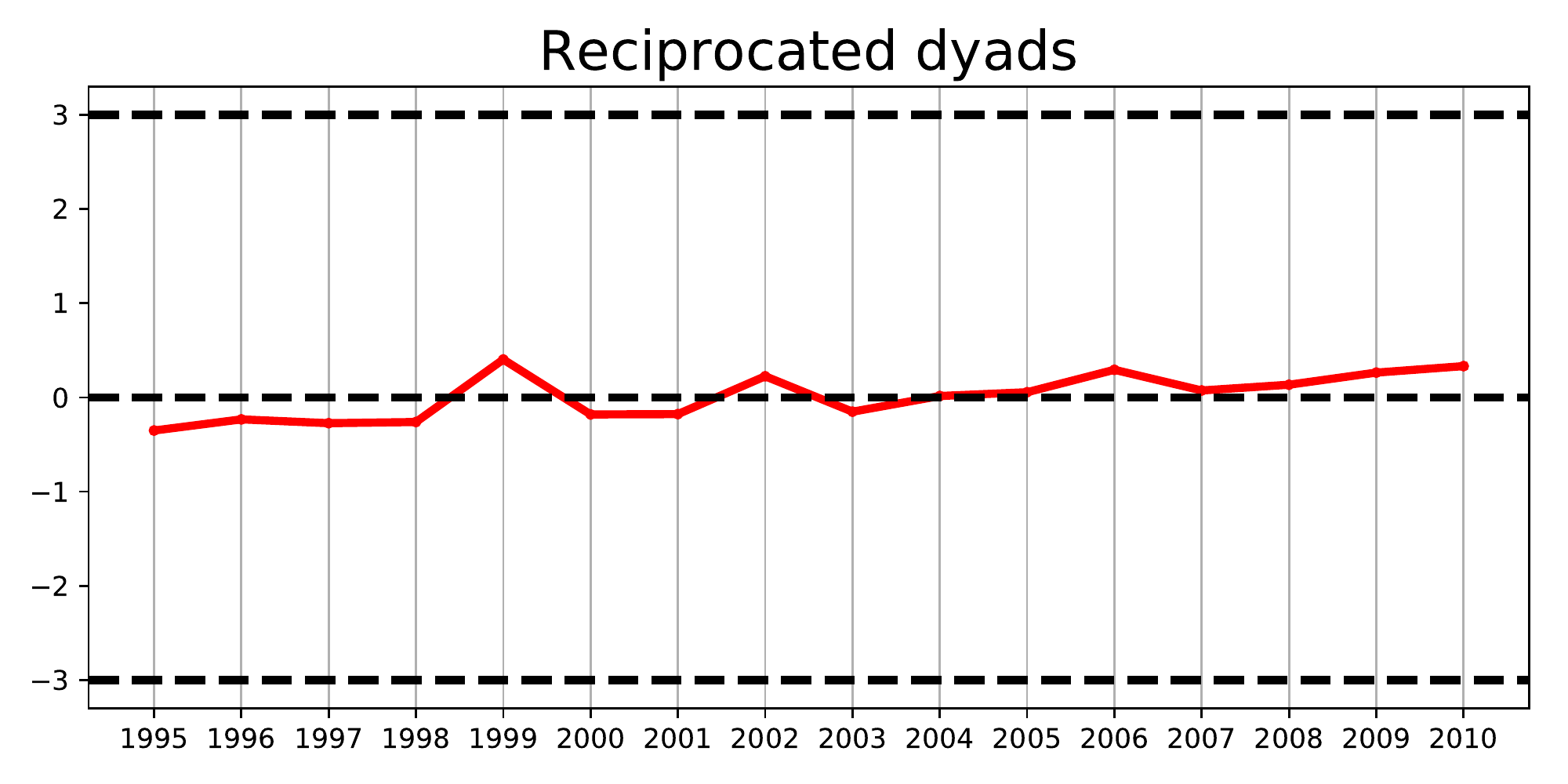}}
    \end{center}
    \caption{Temporal series of z-score statistics for different types of dyadic motifs.}
	\label{fig:dyads}
\end{figure}

As shown in Fig.~\ref{fig:dyads}, we can observe that at the fixed significance level the null model DCM is underestimating the number of reciprocated links (Fig.~\ref{fig:dyads}e), instead the RCM
is able to reproduce almost all the dyadic motifs. From this fact we can assess that reciprocated links play an important role in the topological structure on the network, not captured by the DCM ensemble.

\begin{figure}[h]
	\begin{center}
	\makebox[\textwidth][c]{
    \subfloat[DCM]
	{\includegraphics[scale=0.22]{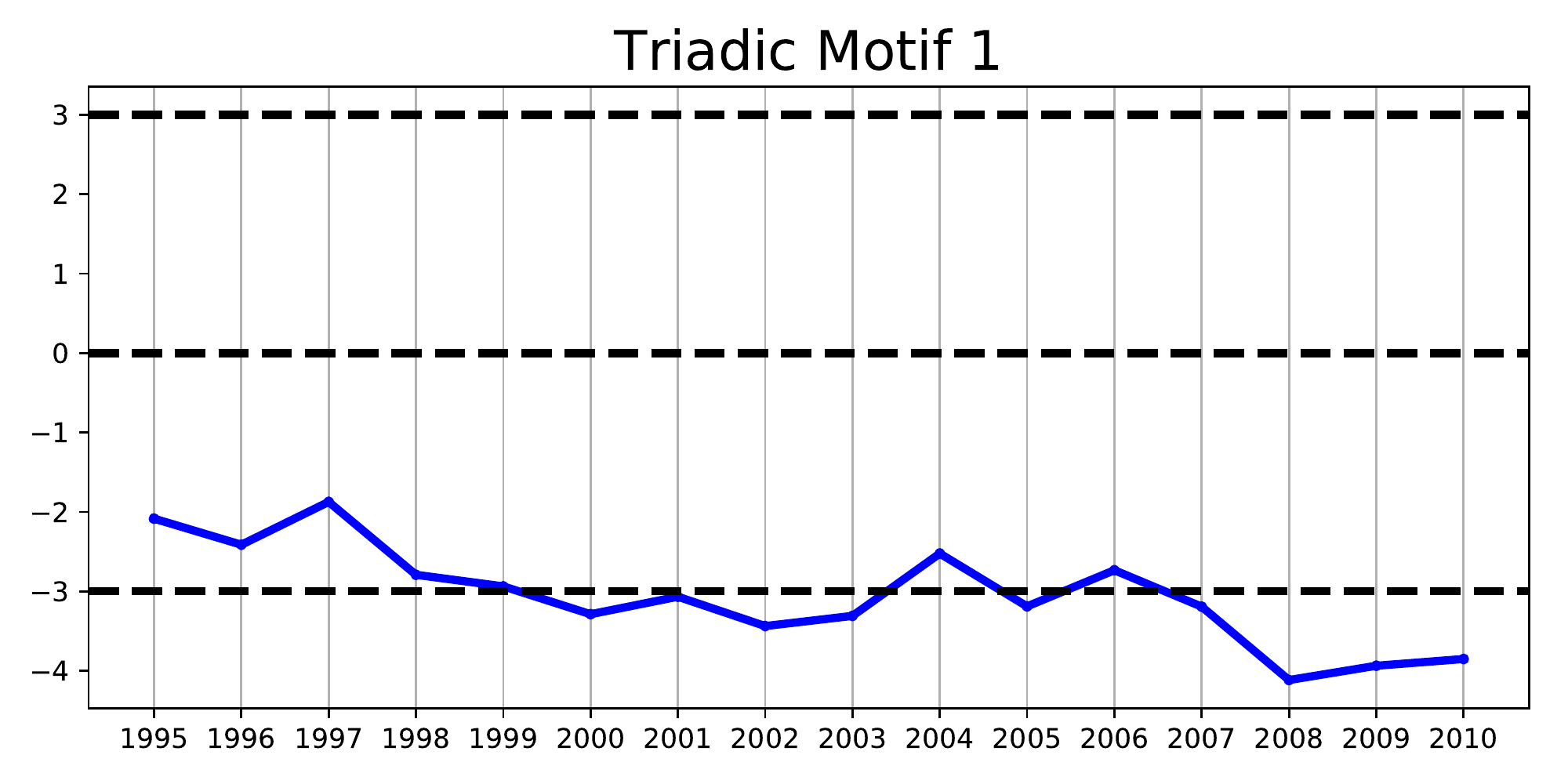}}
    \subfloat[RCM]
    {\includegraphics[scale=0.22]{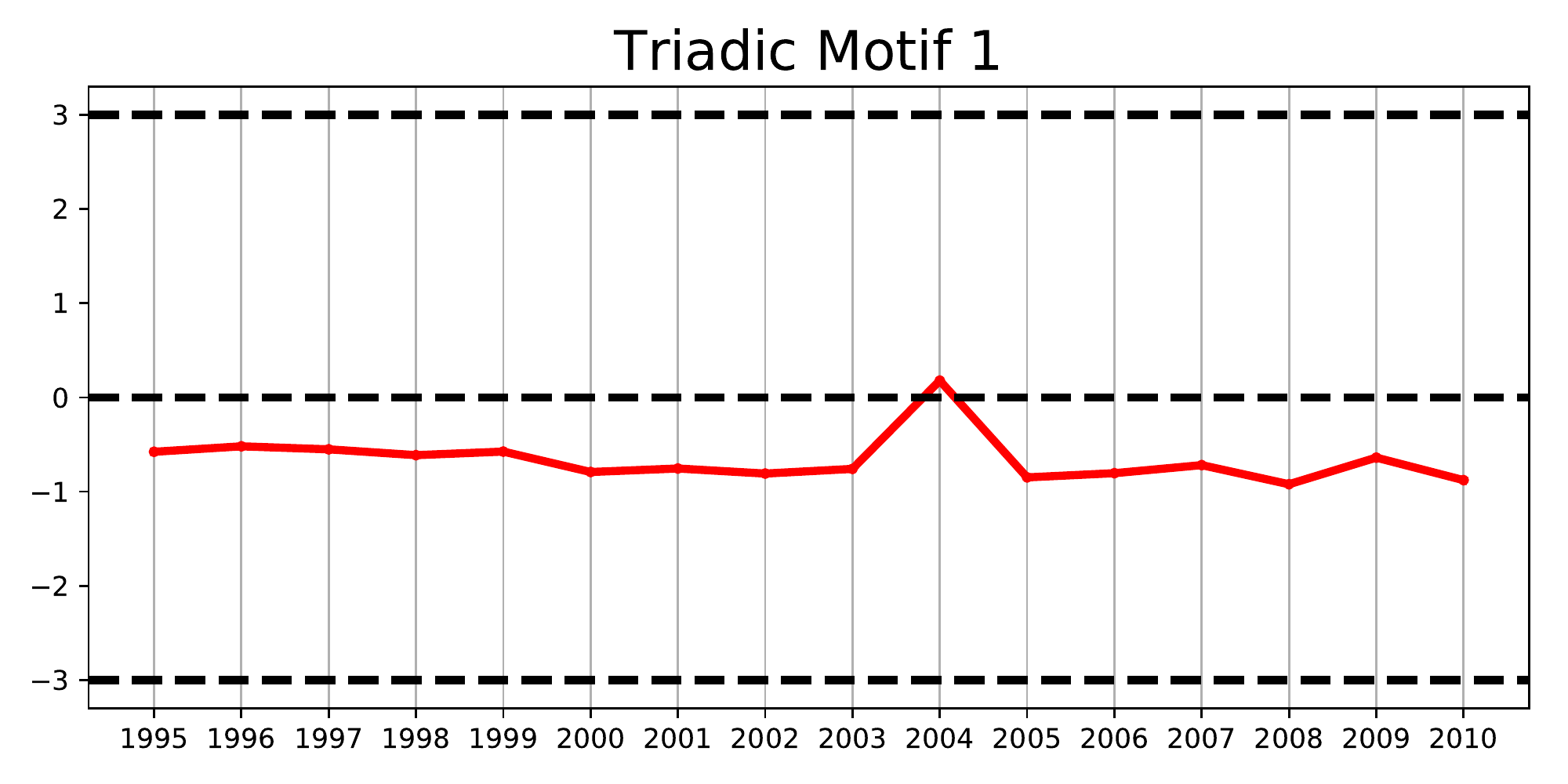}}\qquad
    \subfloat[DCM]
    {\includegraphics[scale=0.22]{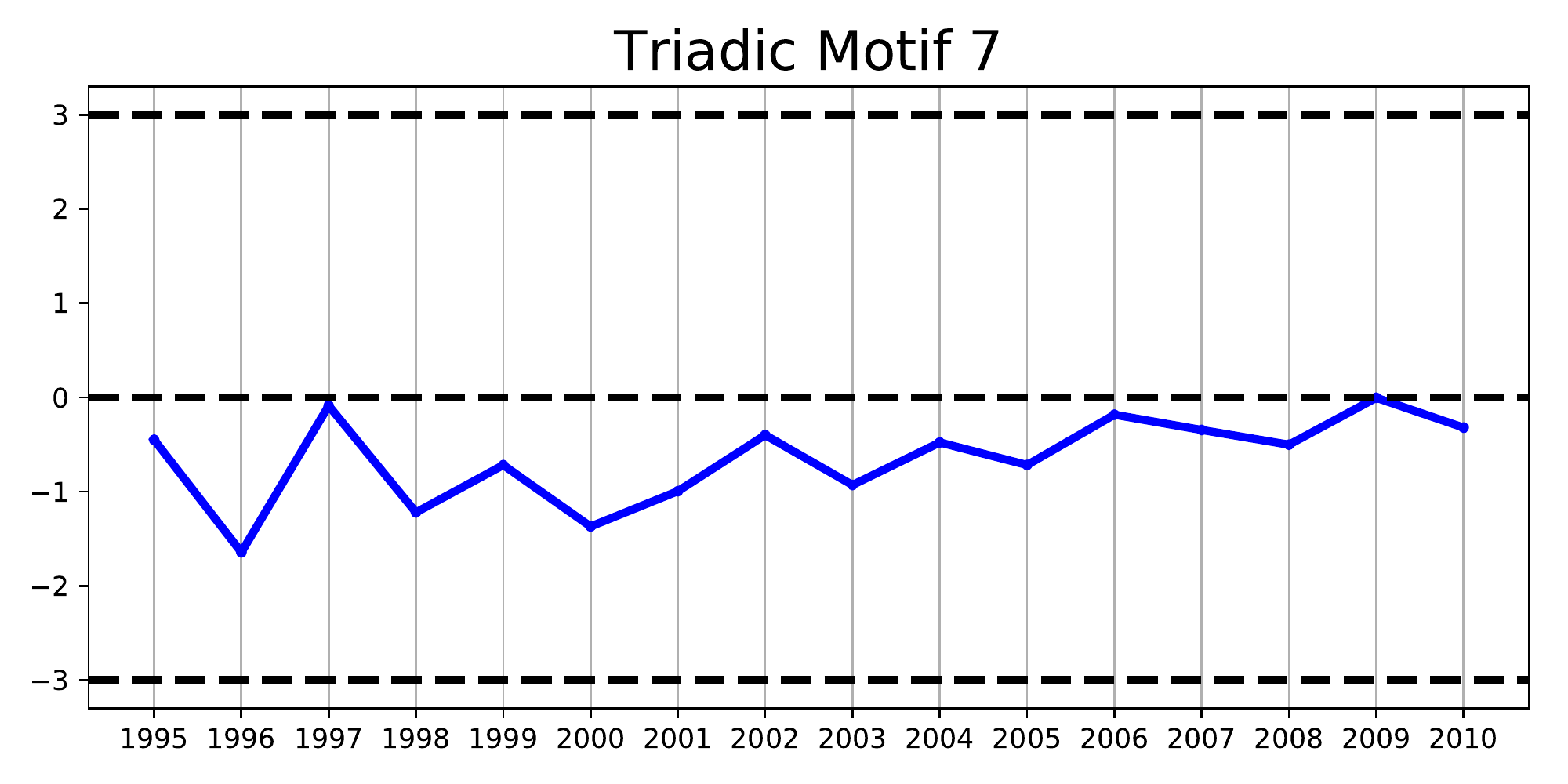}}
    \subfloat[RCM]
    {\includegraphics[scale=0.22]{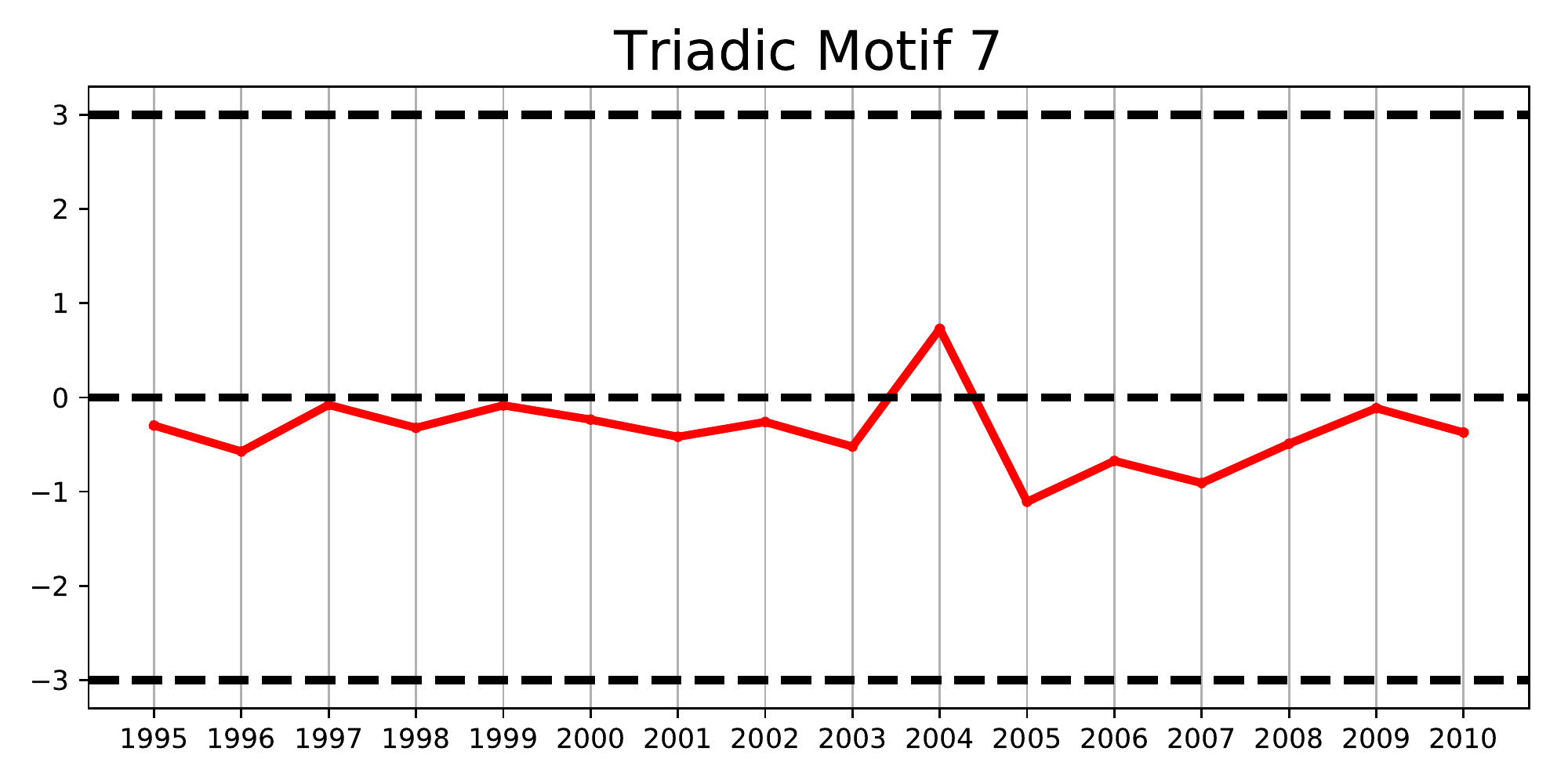}}
    }\\
    \makebox[\textwidth][c]{
    \subfloat[]
	{\includegraphics[scale=0.22]{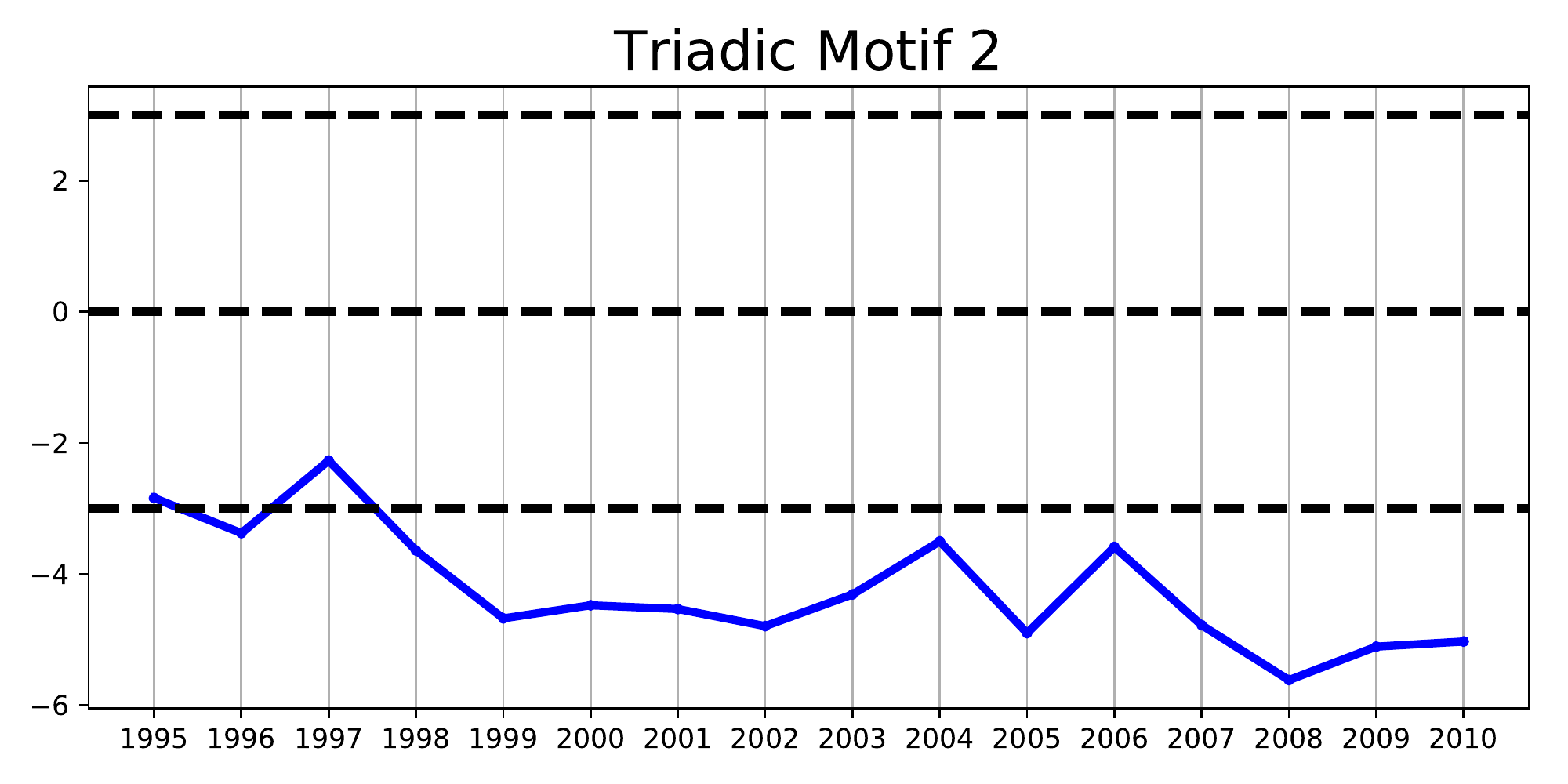}}
    \subfloat[]
    {\includegraphics[scale=0.22]{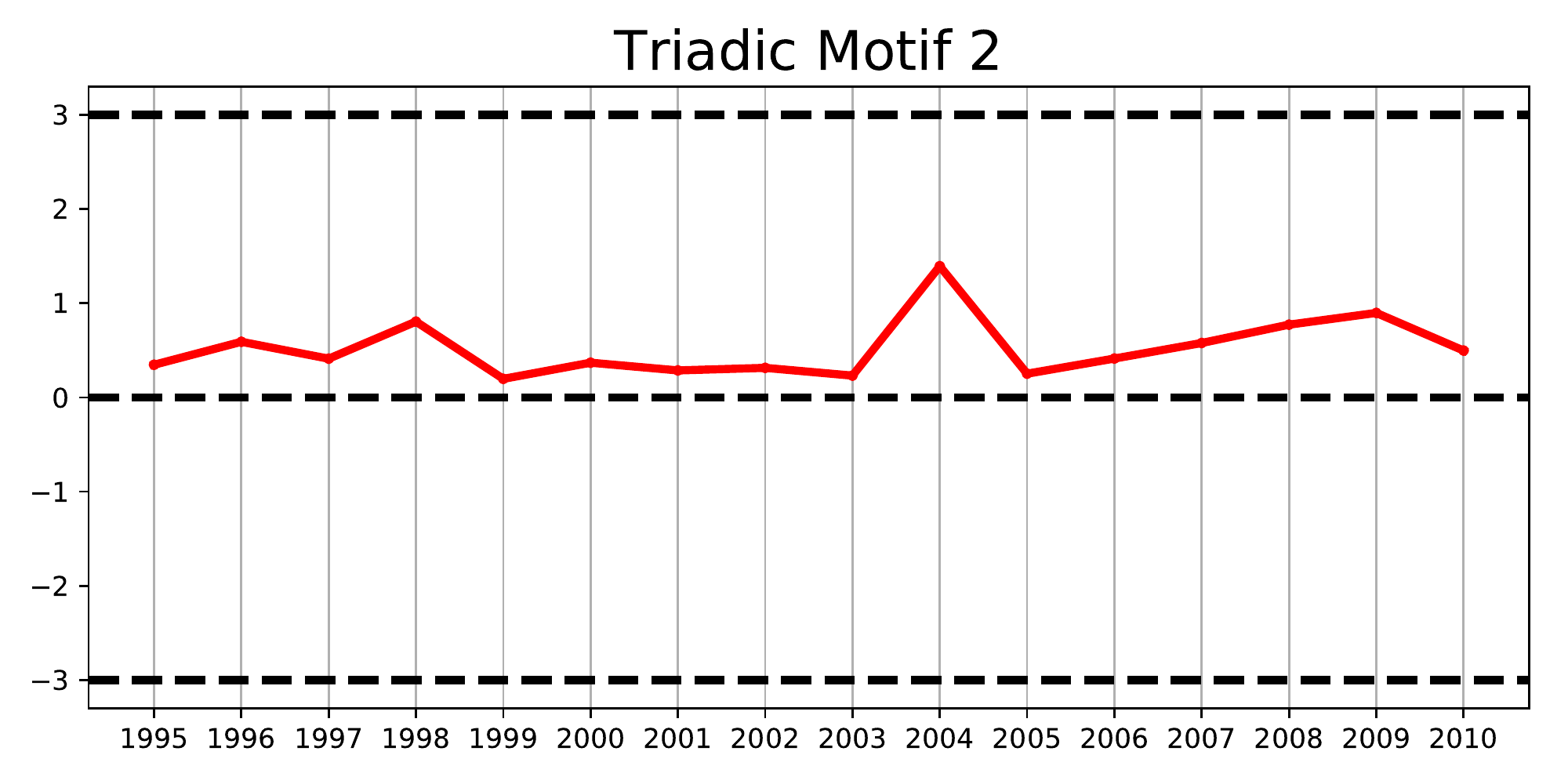}}\qquad
    \subfloat[]
	{\includegraphics[scale=0.22]{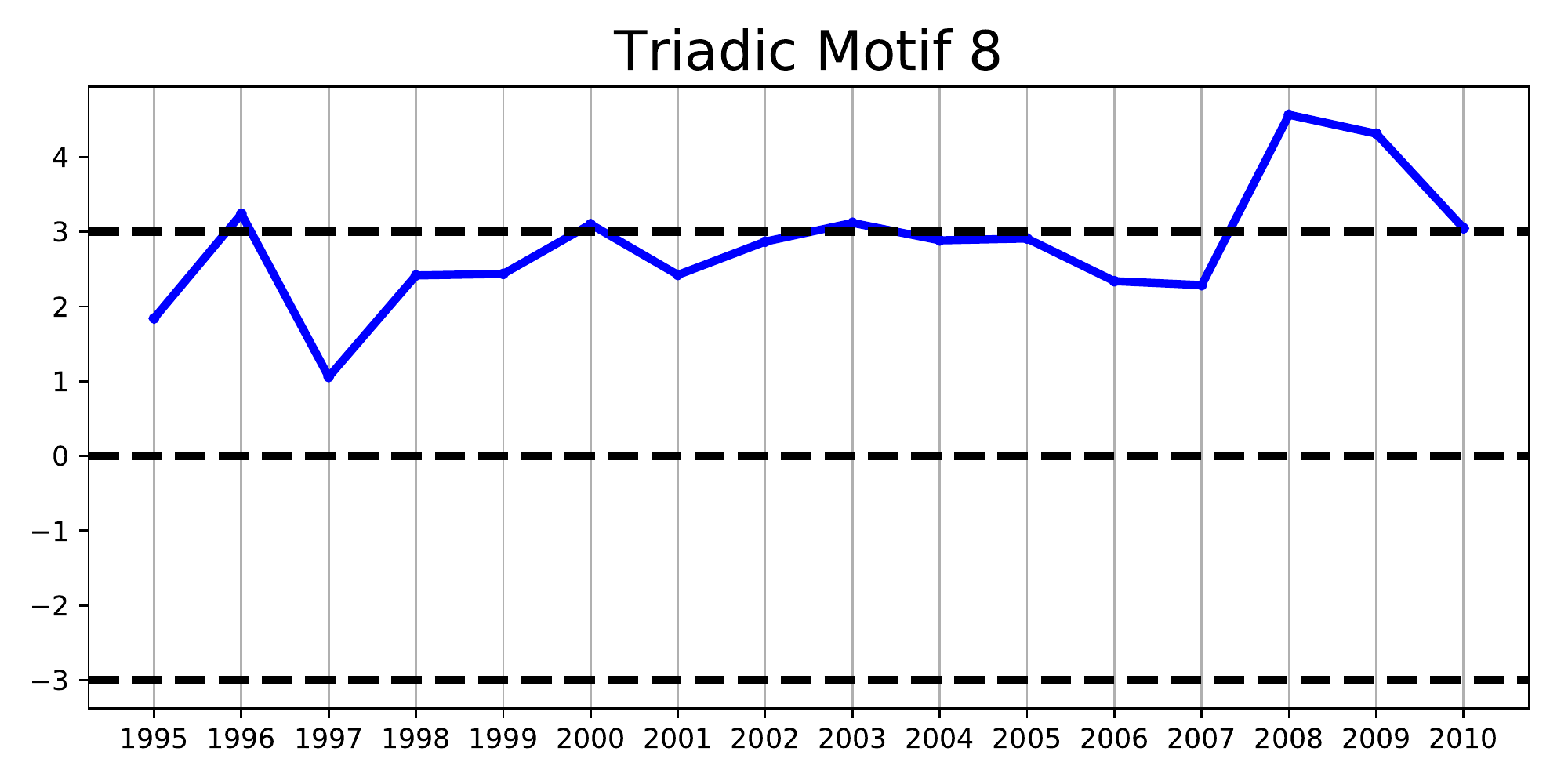}}
    \subfloat[]
    {\includegraphics[scale=0.22]{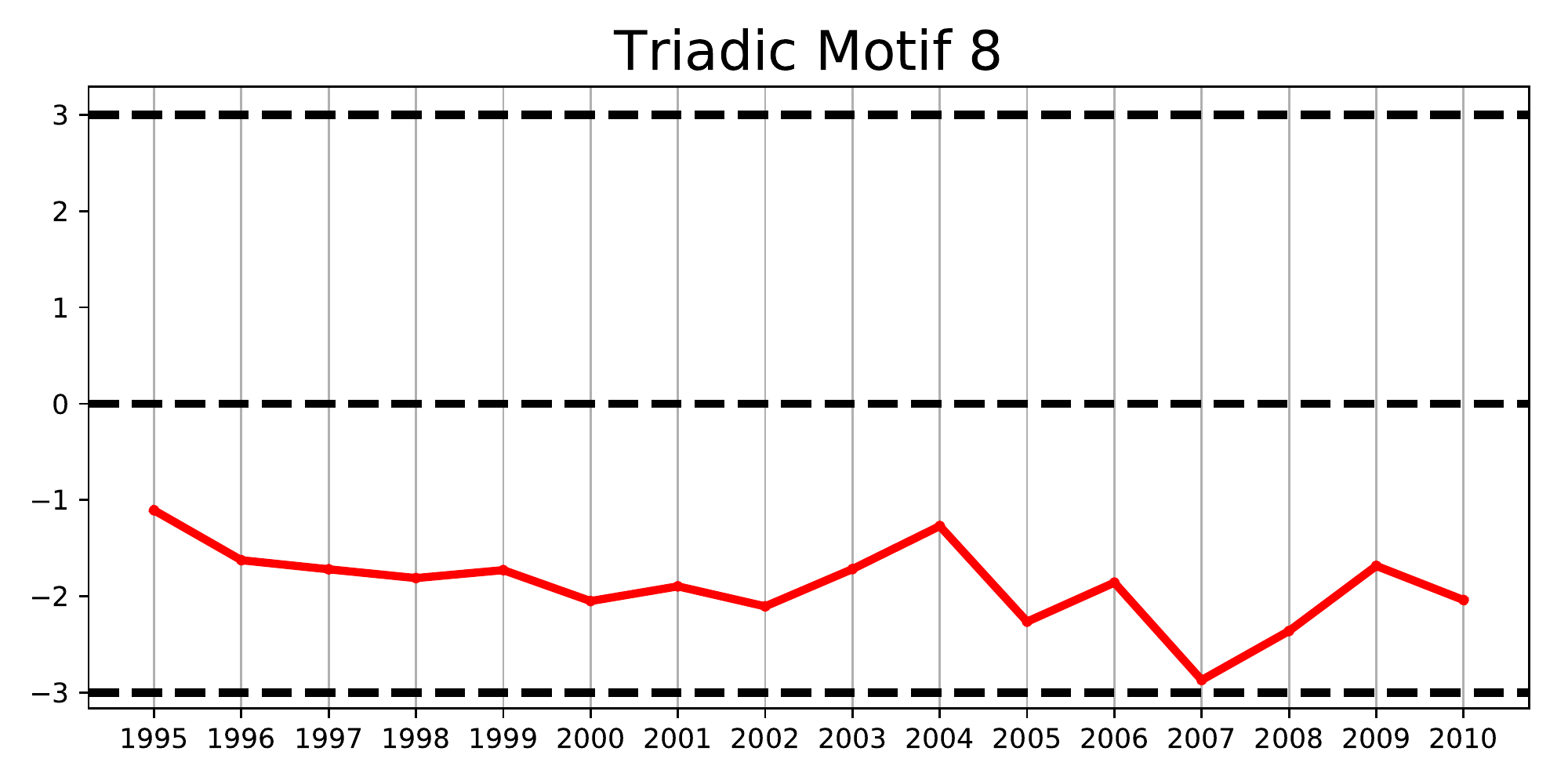}}
    }
    \makebox[\textwidth][c]{
    \subfloat[]
	{\includegraphics[scale=0.22]{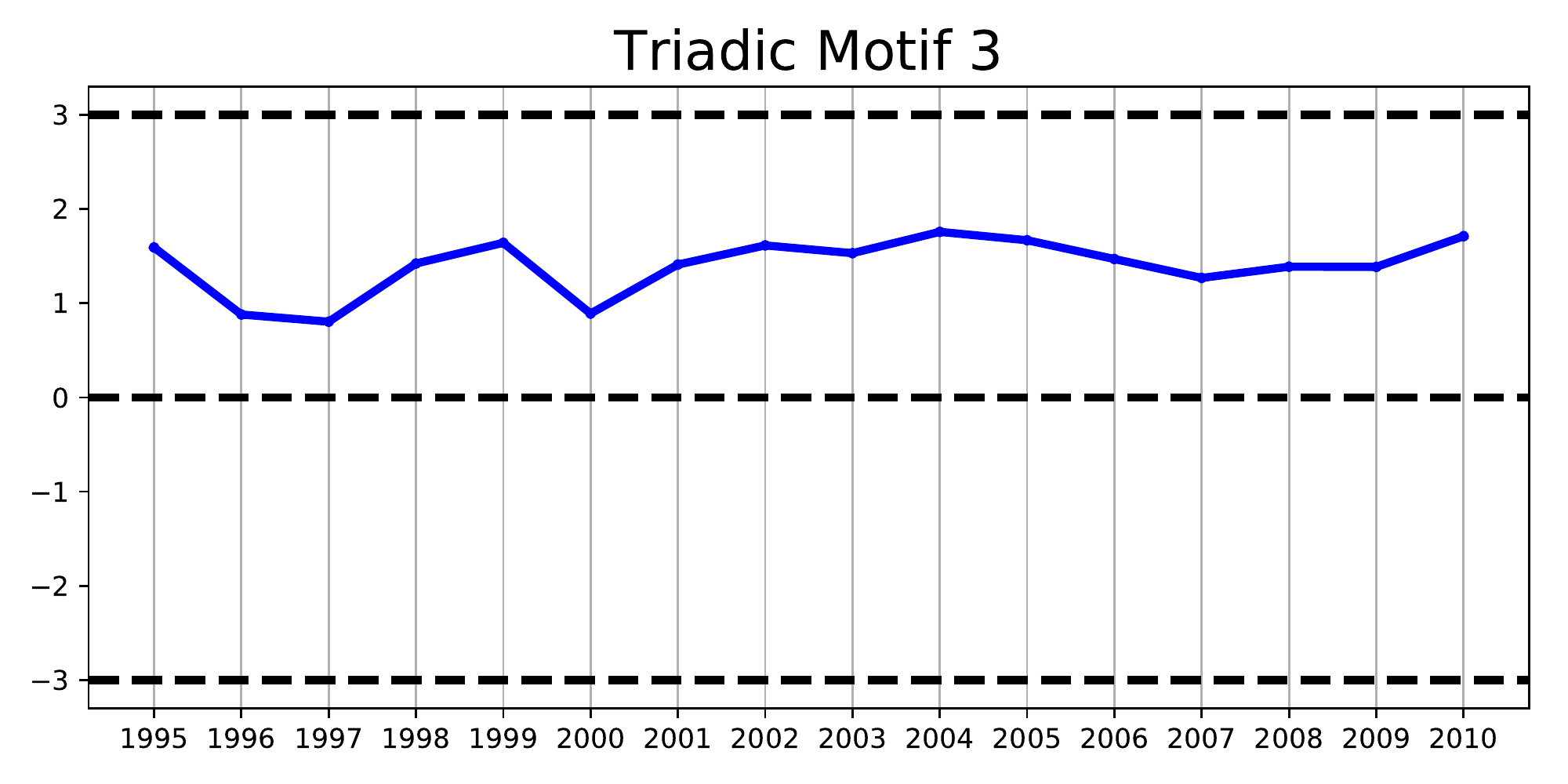}}
    \subfloat[]
    {\includegraphics[scale=0.22]{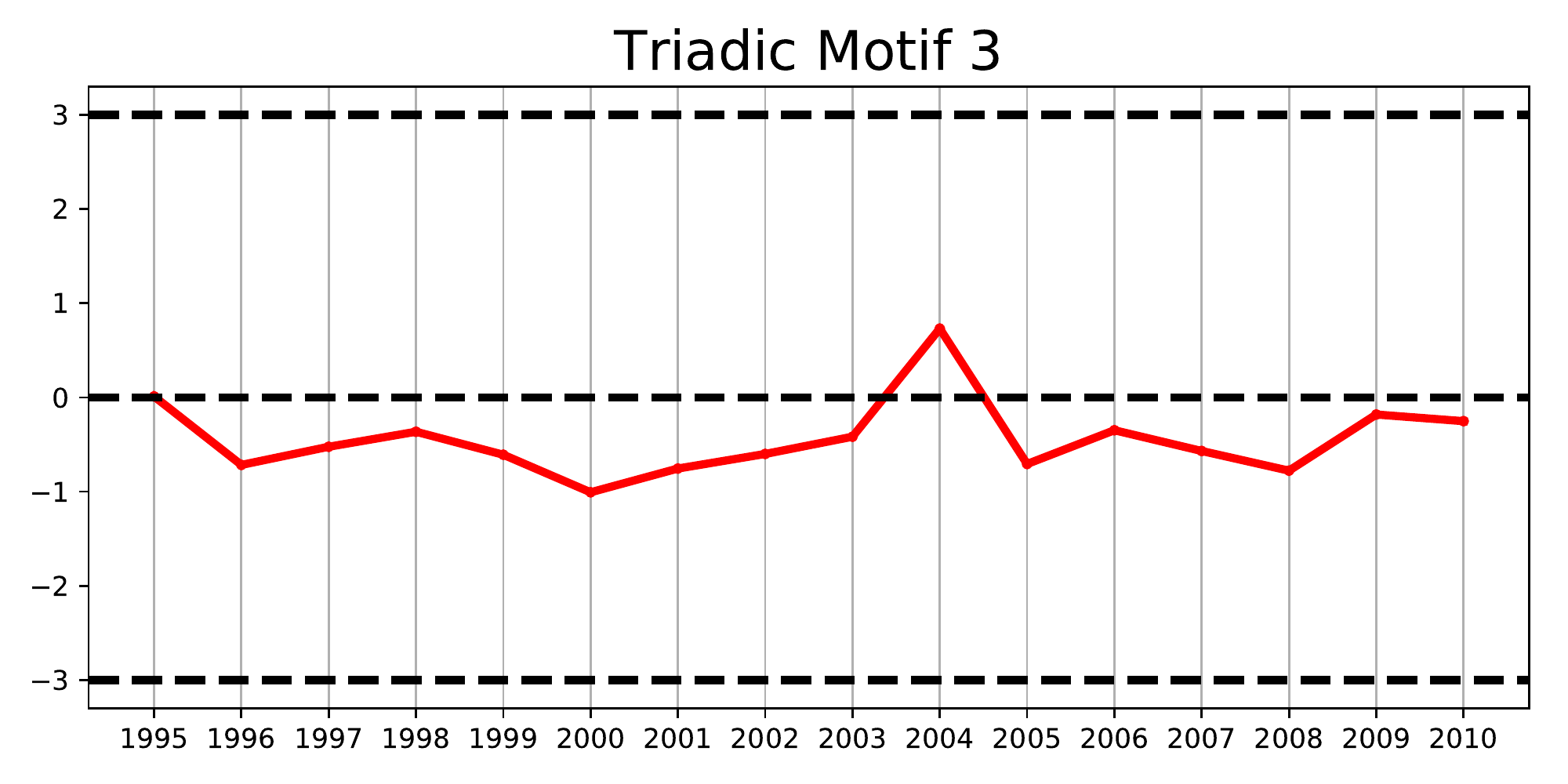}}\qquad
    \subfloat[]
	{\includegraphics[scale=0.22]{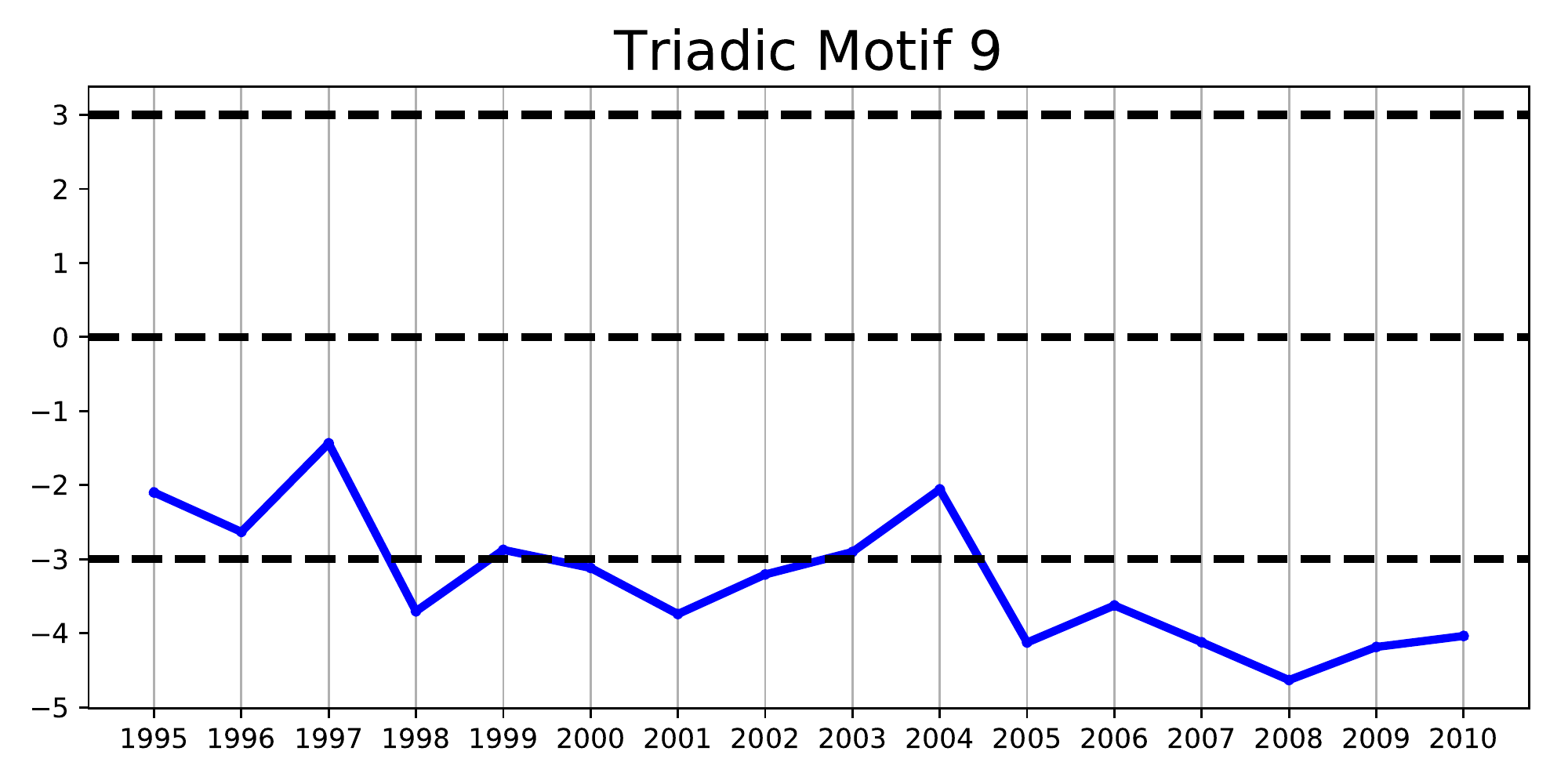}}
    \subfloat[]
    {\includegraphics[scale=0.22]{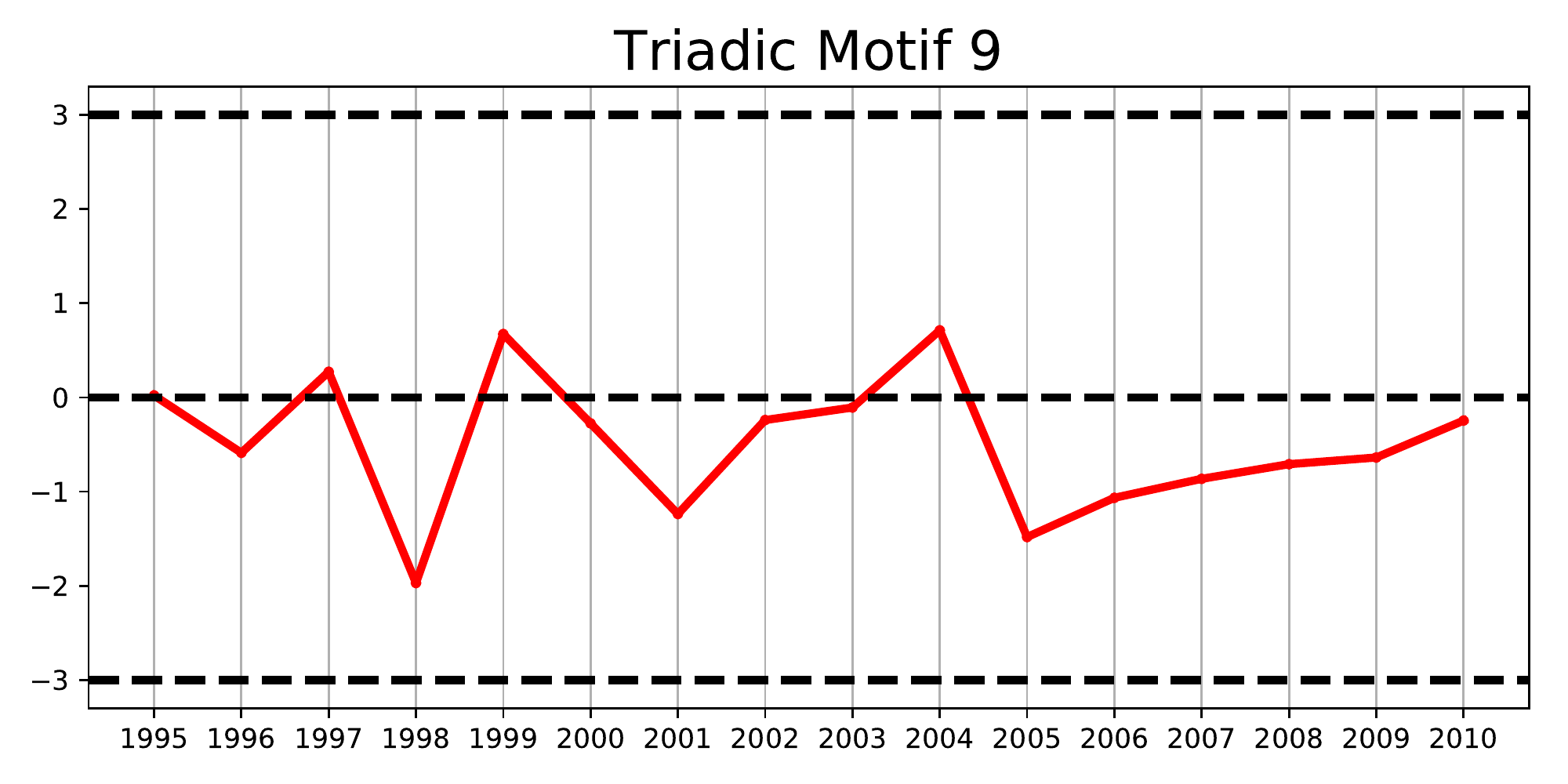}}
    }
    \makebox[\textwidth][c]{
    \subfloat[]
	{\includegraphics[scale=0.22]{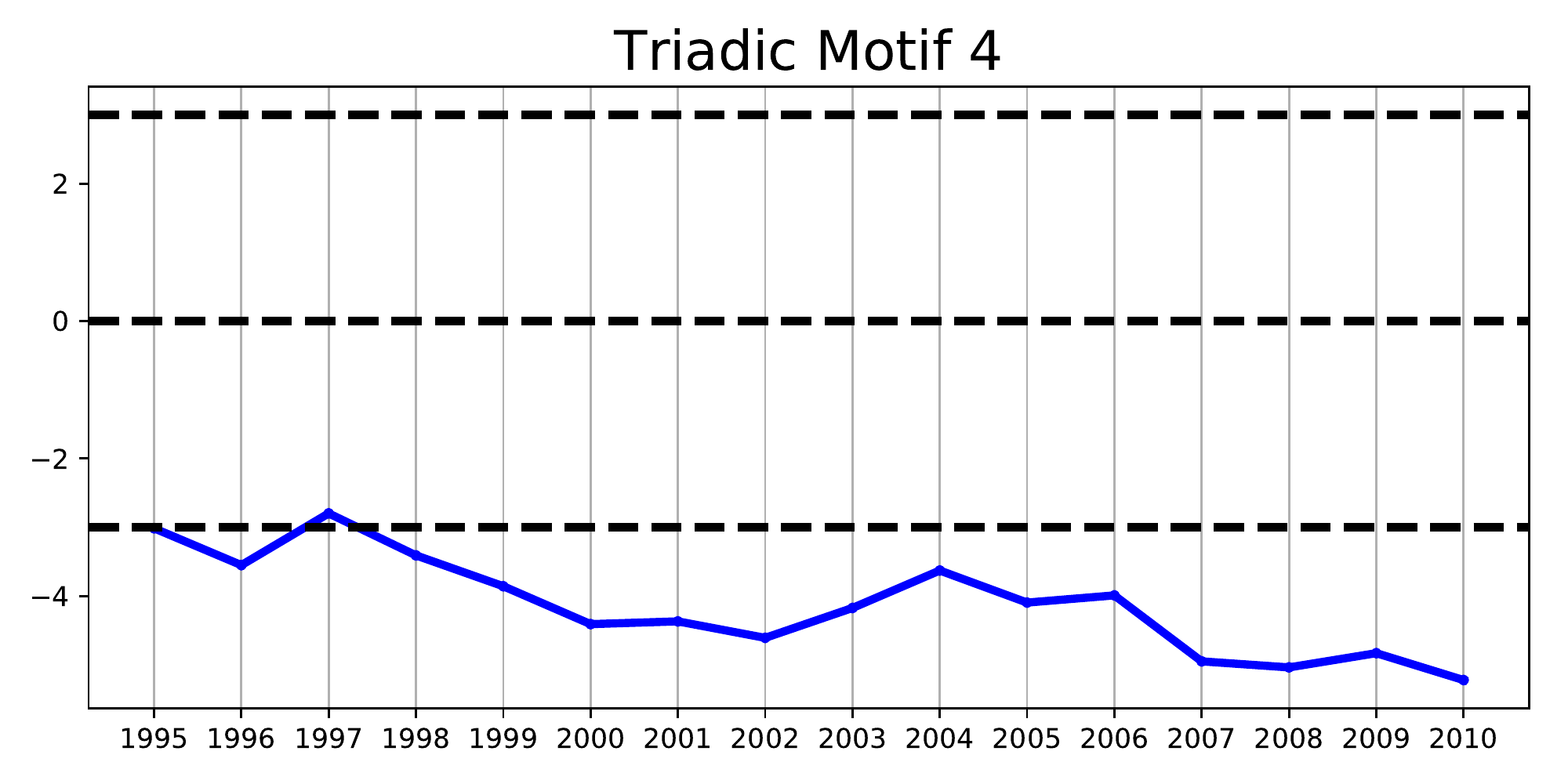}}
    \subfloat[]
    {\includegraphics[scale=0.22]{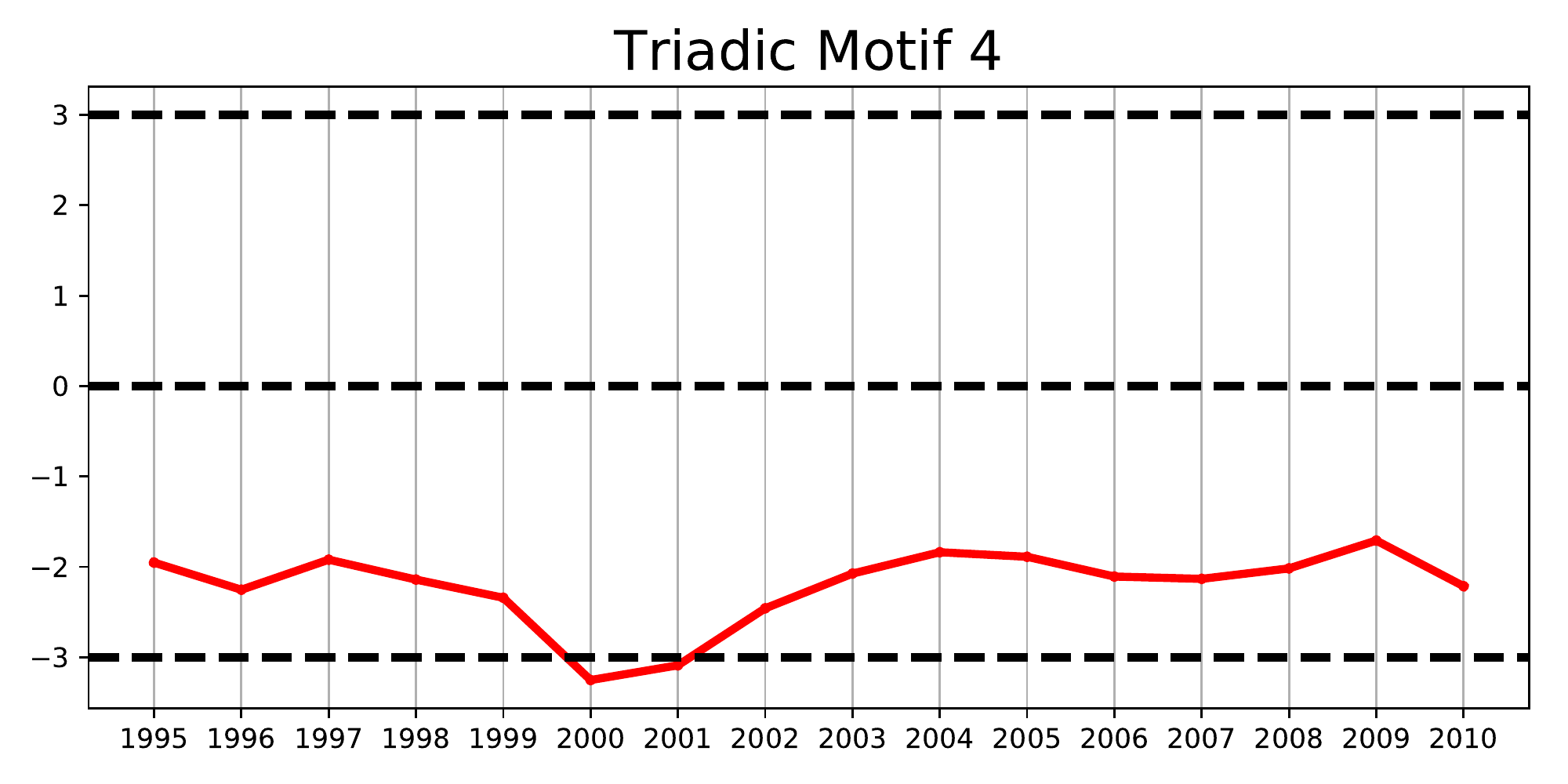}}\qquad
    \subfloat[]
	{\includegraphics[scale=0.22]{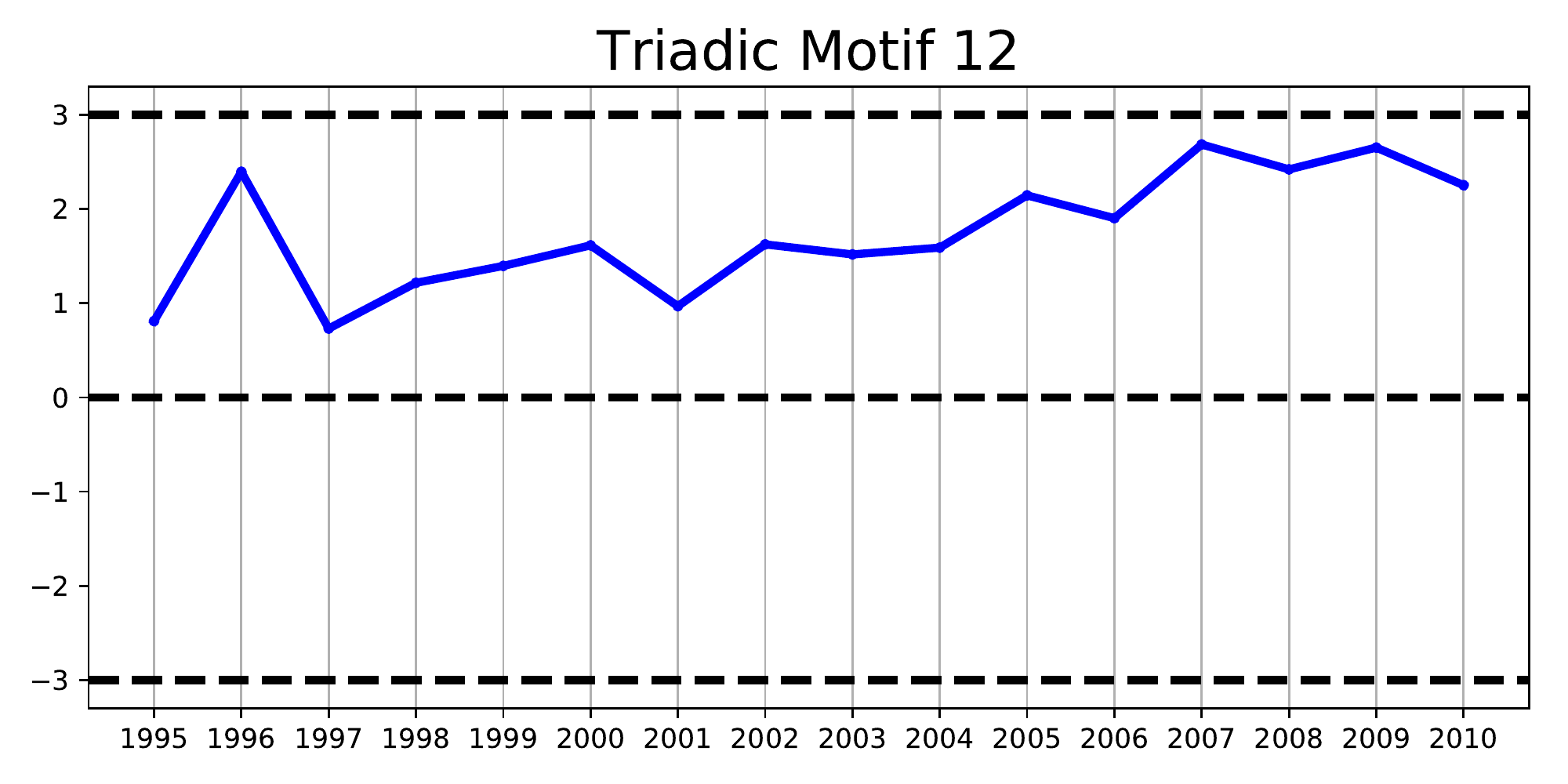}}
    \subfloat[]
    {\includegraphics[scale=0.22]{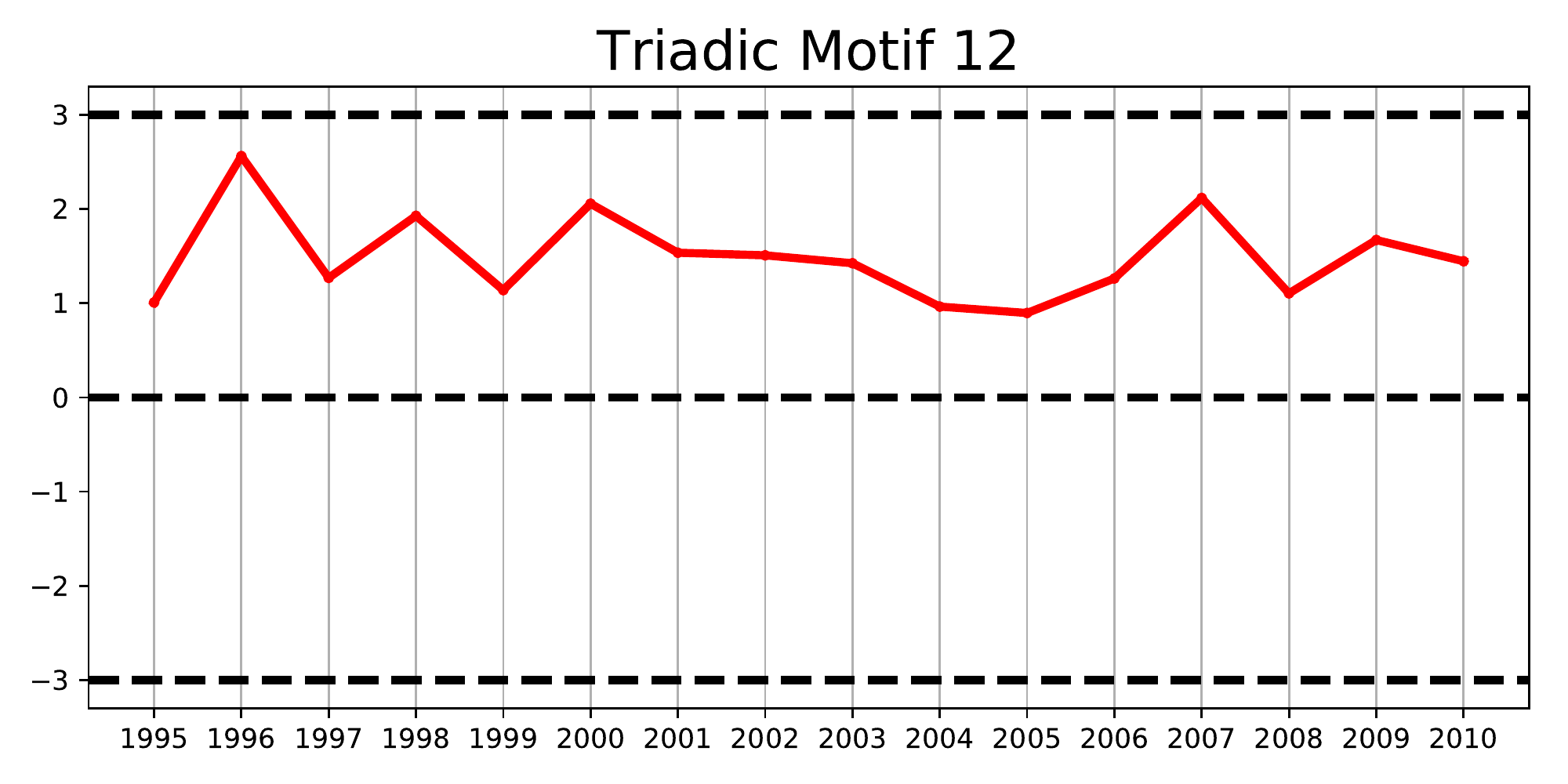}}
    }
    \makebox[\textwidth][c]{
    \subfloat[]
	{\includegraphics[scale=0.22]{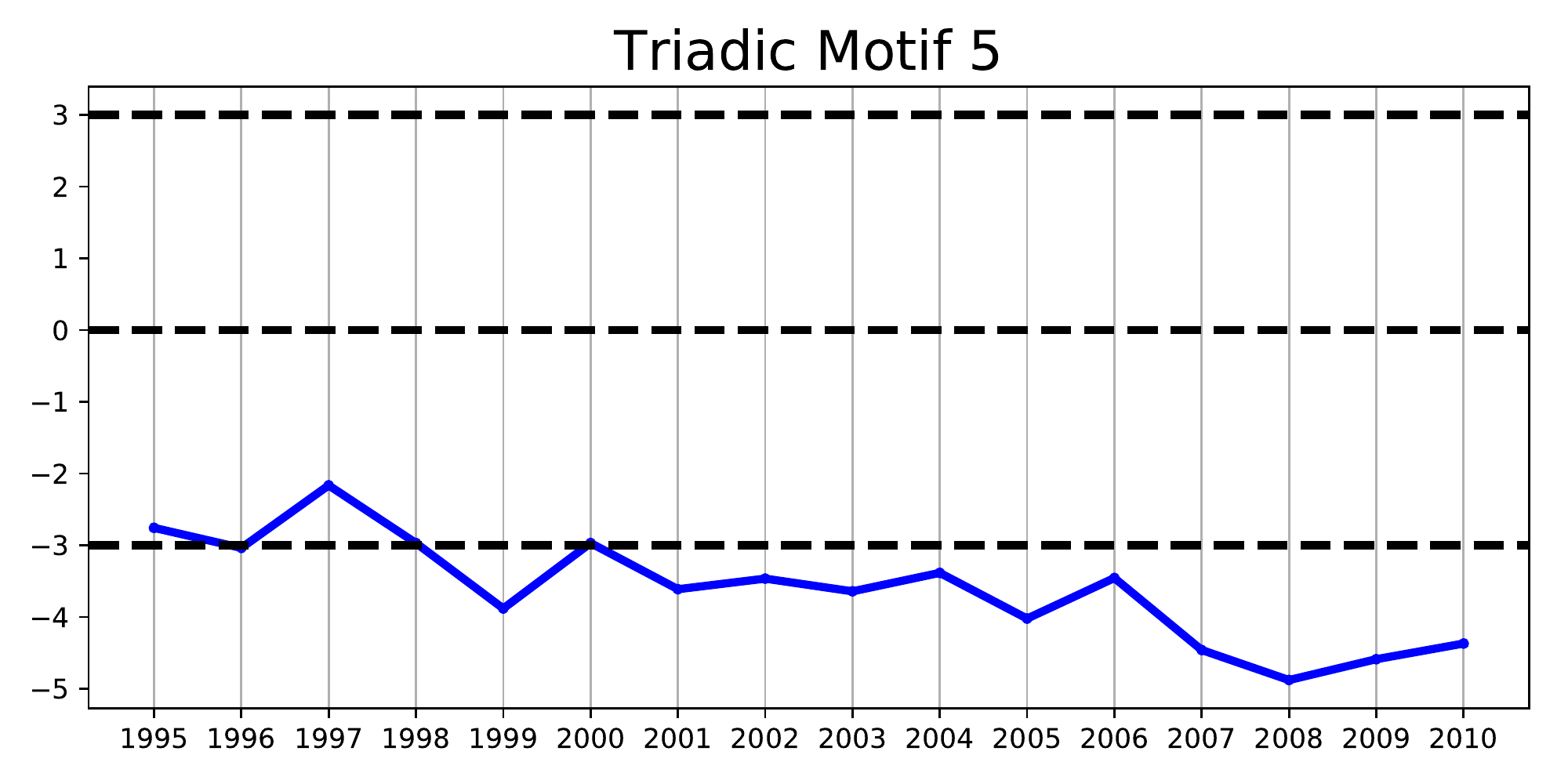}}
    \subfloat[]
    {\includegraphics[scale=0.22]{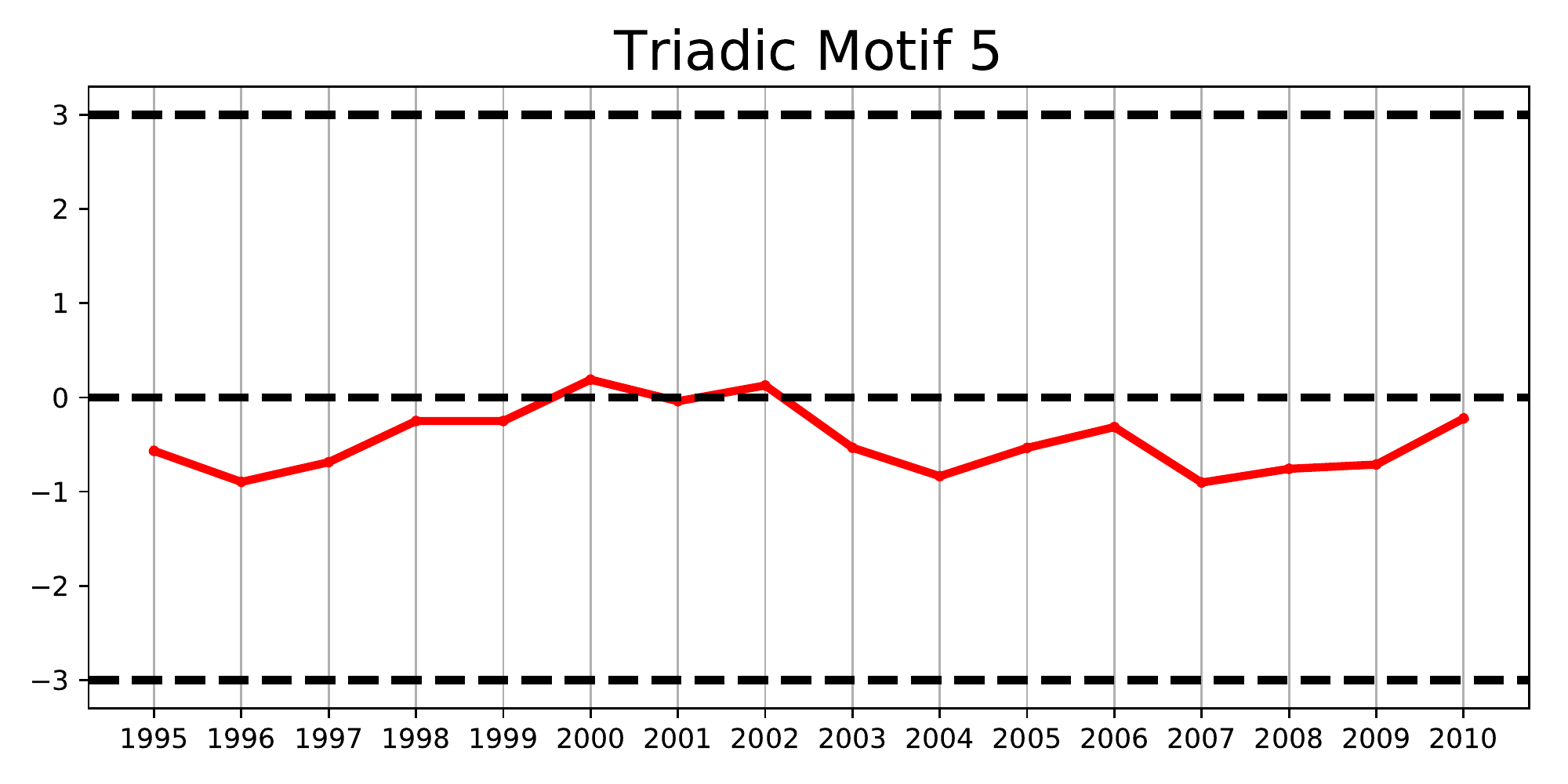}}\qquad
    \subfloat[]
	{\includegraphics[scale=0.22]{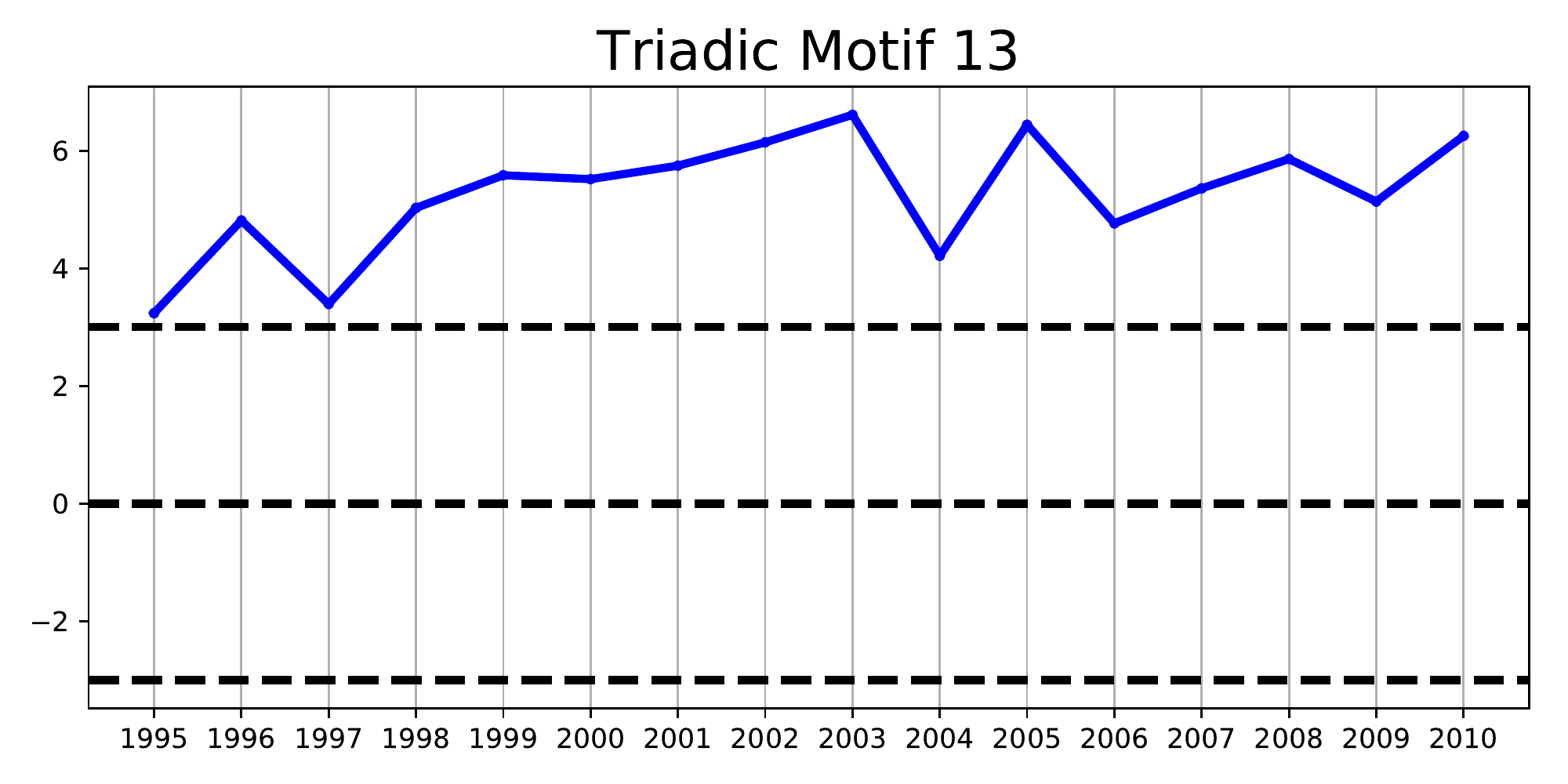}}
    \subfloat[]
    {\includegraphics[scale=0.22]{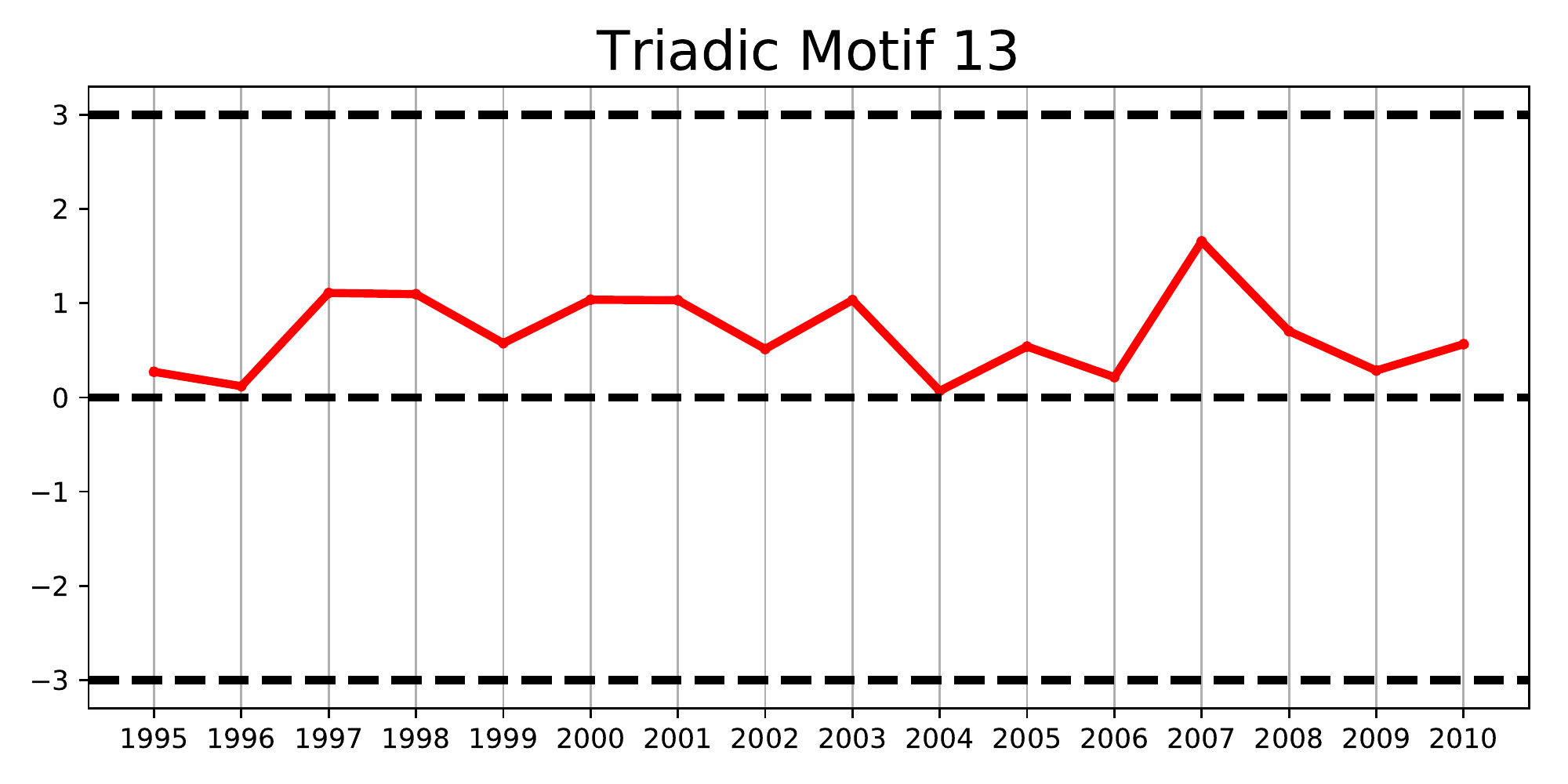}}
    }
    \end{center}
    \caption{Temporal series of z-score statistics for several types of triadic motifs in the binary case.}
	\label{fig:triads}
\end{figure}

To deep understand the presence of higher order motifs we have analyzed also the statistics of the triadic ones, starting from the binary models, whose temporal series are reported in Fig.~\ref{fig:triads}. As in the dyadic case the RCM is able to reproduce the original dataset in almost every motif. From the DCM side we can see instead that some motifs are not well reproduced during different snapshots. Motifs 8-13 are in average underestimated by the null partially or totally over years. Since this motifs are based on reciprocated links, it turns out that also for triadic patterns the presence of reciprocity can not be captured by randomization. Motifs 1-2-4-5-9 are in average overestimated by the null model over the years, plus they have a decreasing trend. This motifs are characterized by complete absence of reciprocity, so this effect is complementary respect what happens for motifs 8-13. It is interesting to notice that motif 8 which was underestimated from the DCM, appears overestimated by the RCM. This sheds light on the necessity to introduce a novel constraint between the node degree and the node number of reciprocated links.

\begin{figure}[h]
	\begin{center}
	\makebox[\textwidth][c]{
    \subfloat[]
    {\includegraphics[scale=0.3]{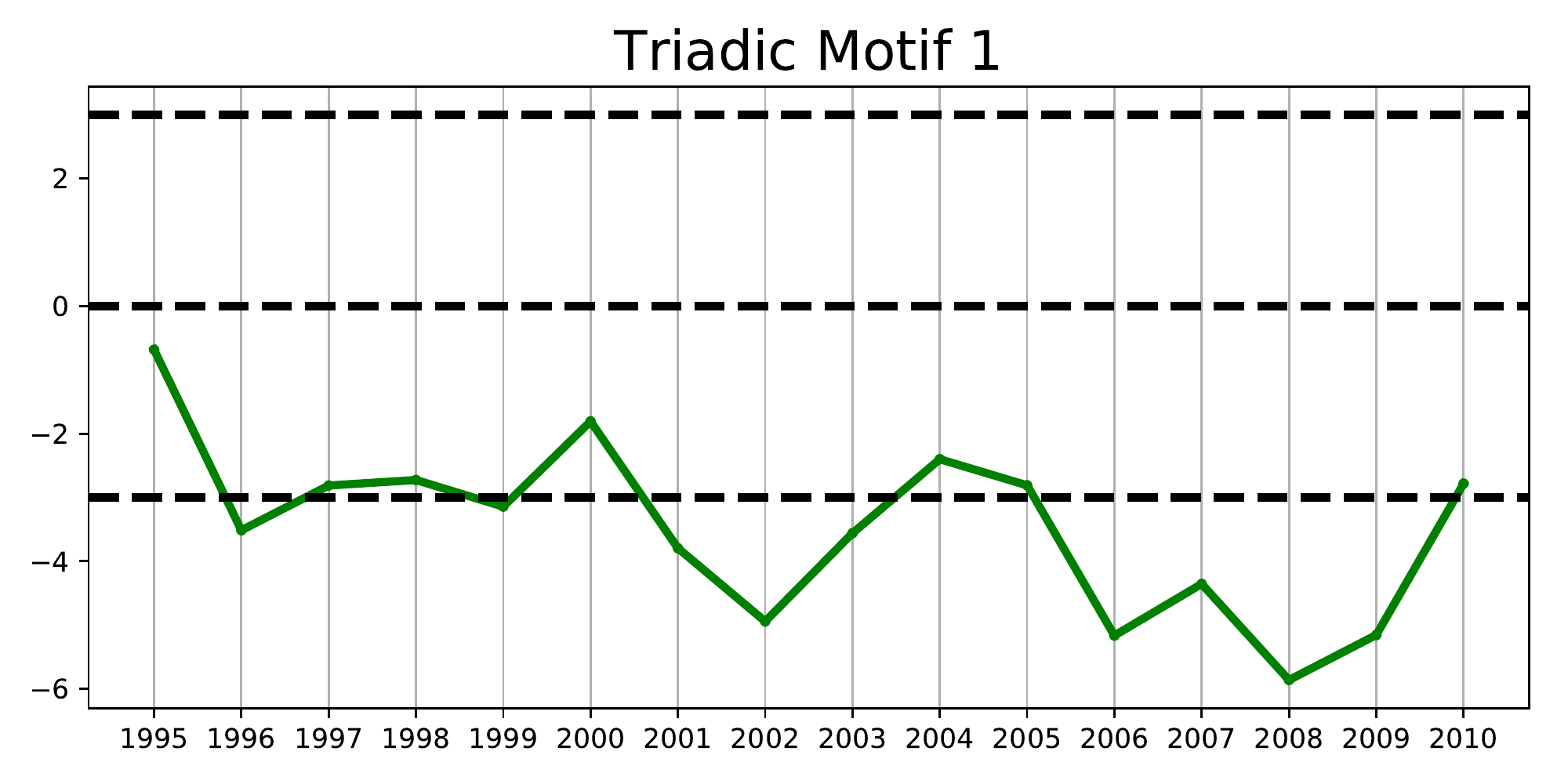}}
    \subfloat[]
	{\includegraphics[scale=0.3]{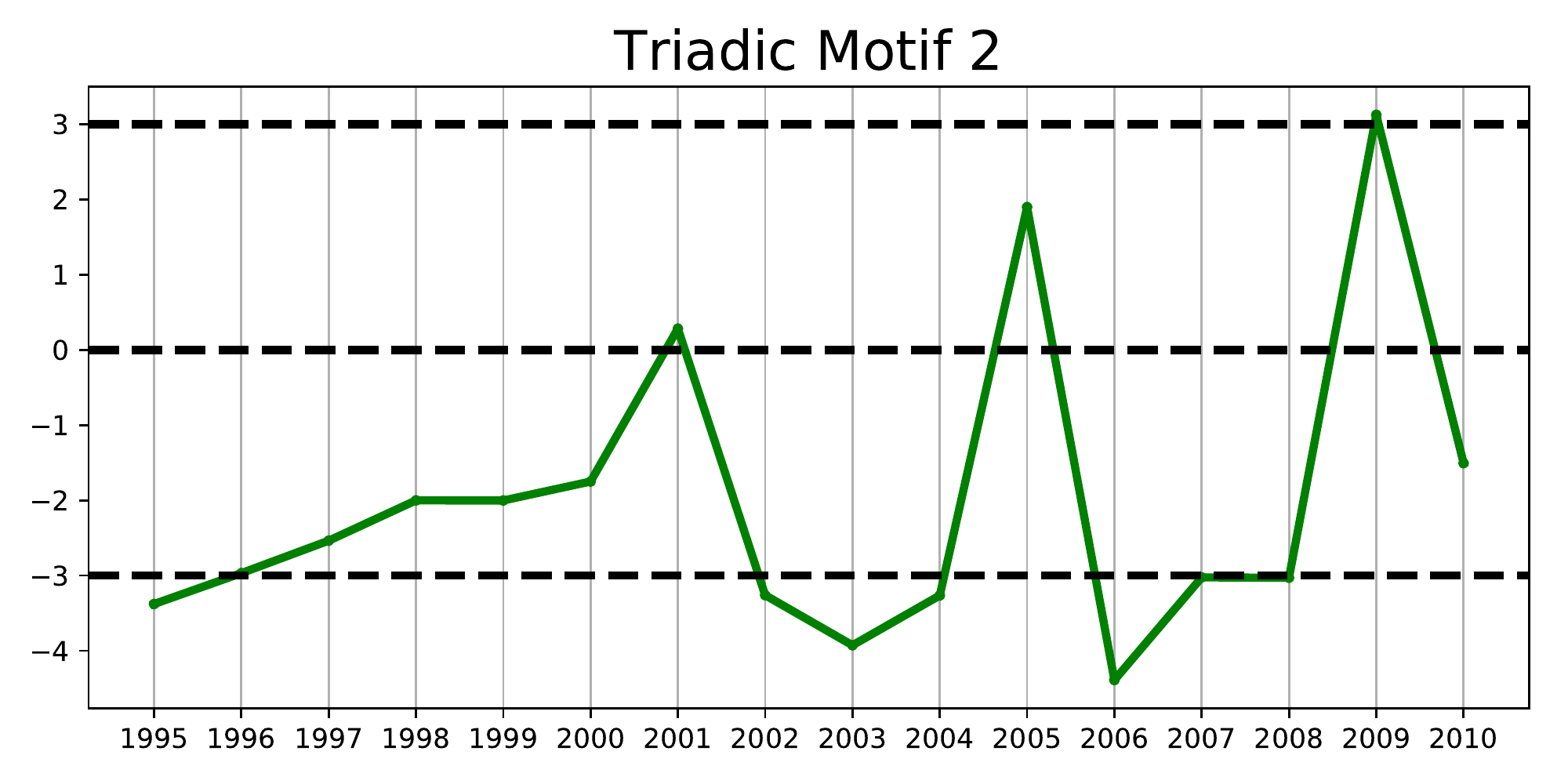}}
	}
    \makebox[\textwidth][c]{
    \subfloat[]
    {\includegraphics[scale=0.3]{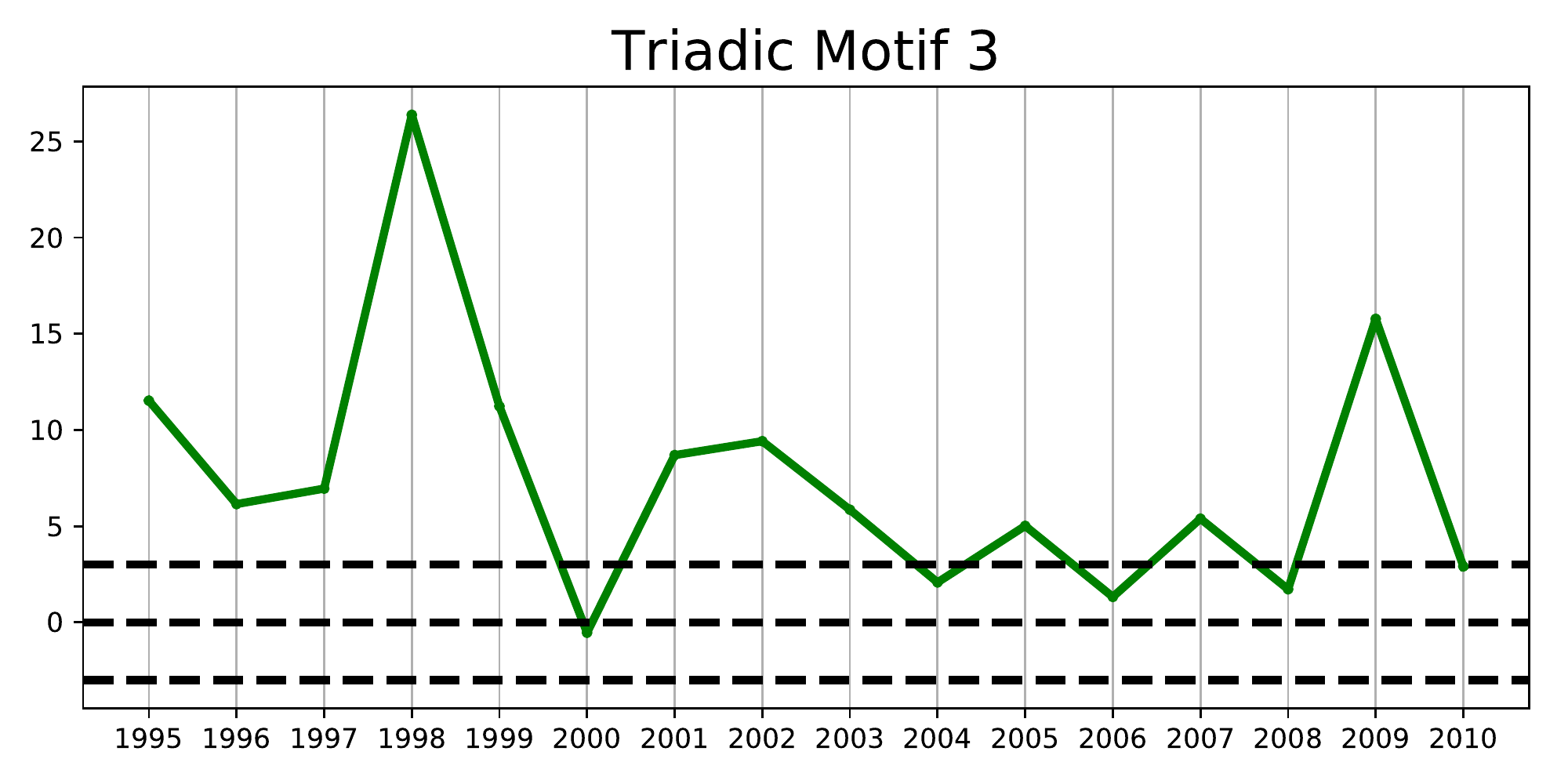}}
    \subfloat[]
	{\includegraphics[scale=0.3]{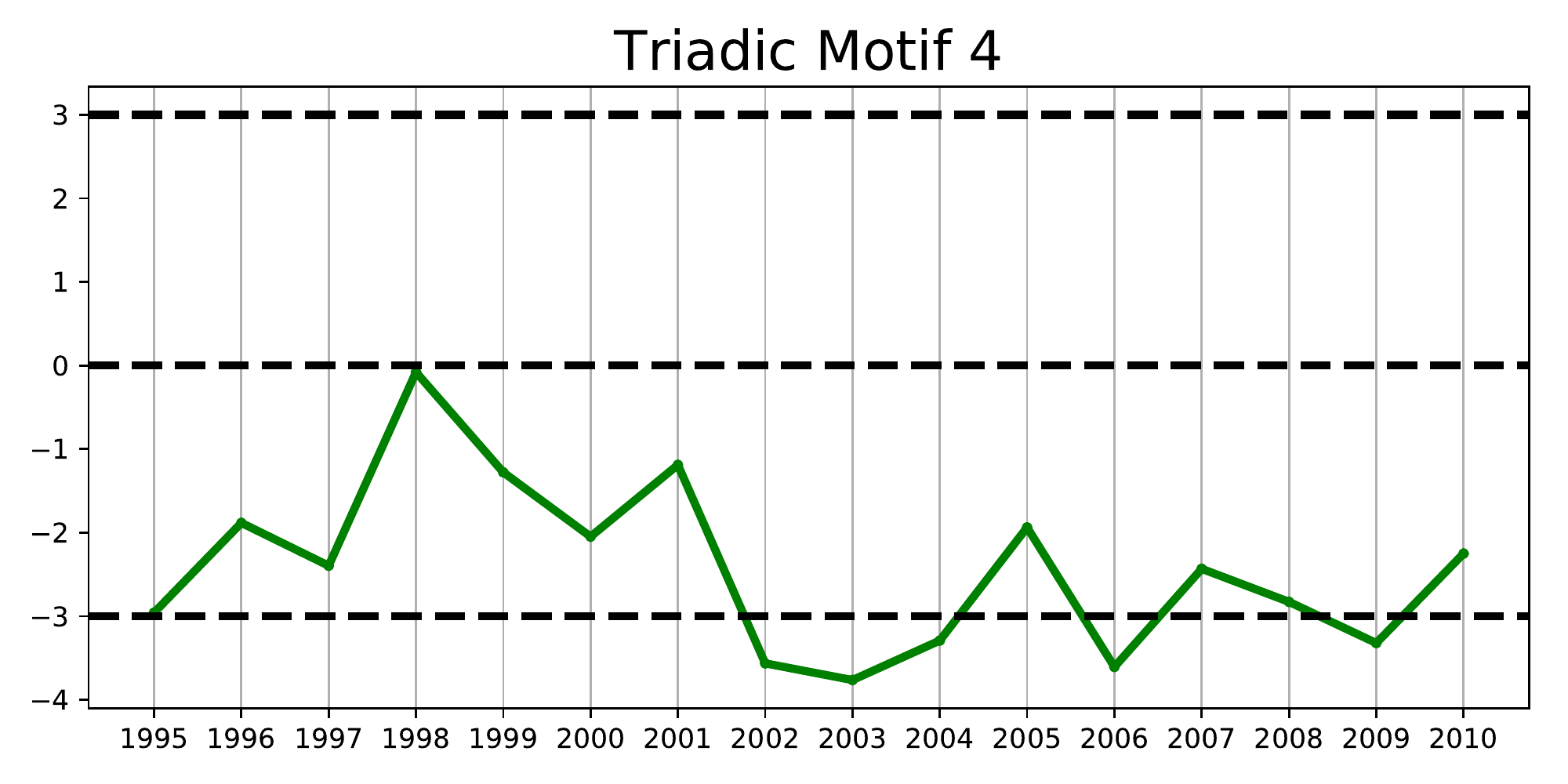}}
	}
	\makebox[\textwidth][c]{
    \subfloat[]
    {\includegraphics[scale=0.3]{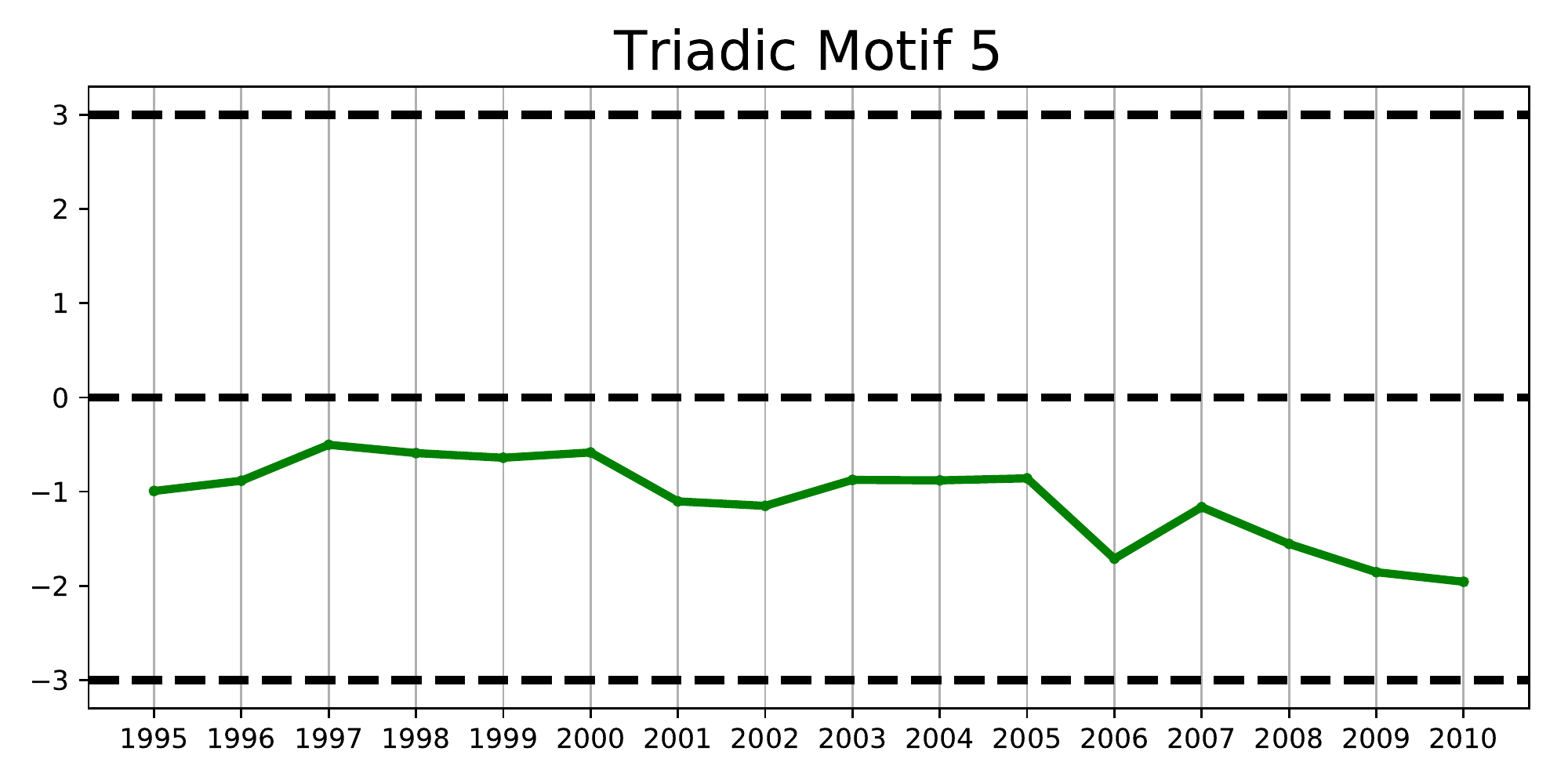}}
    \subfloat[]
	{\includegraphics[scale=0.3]{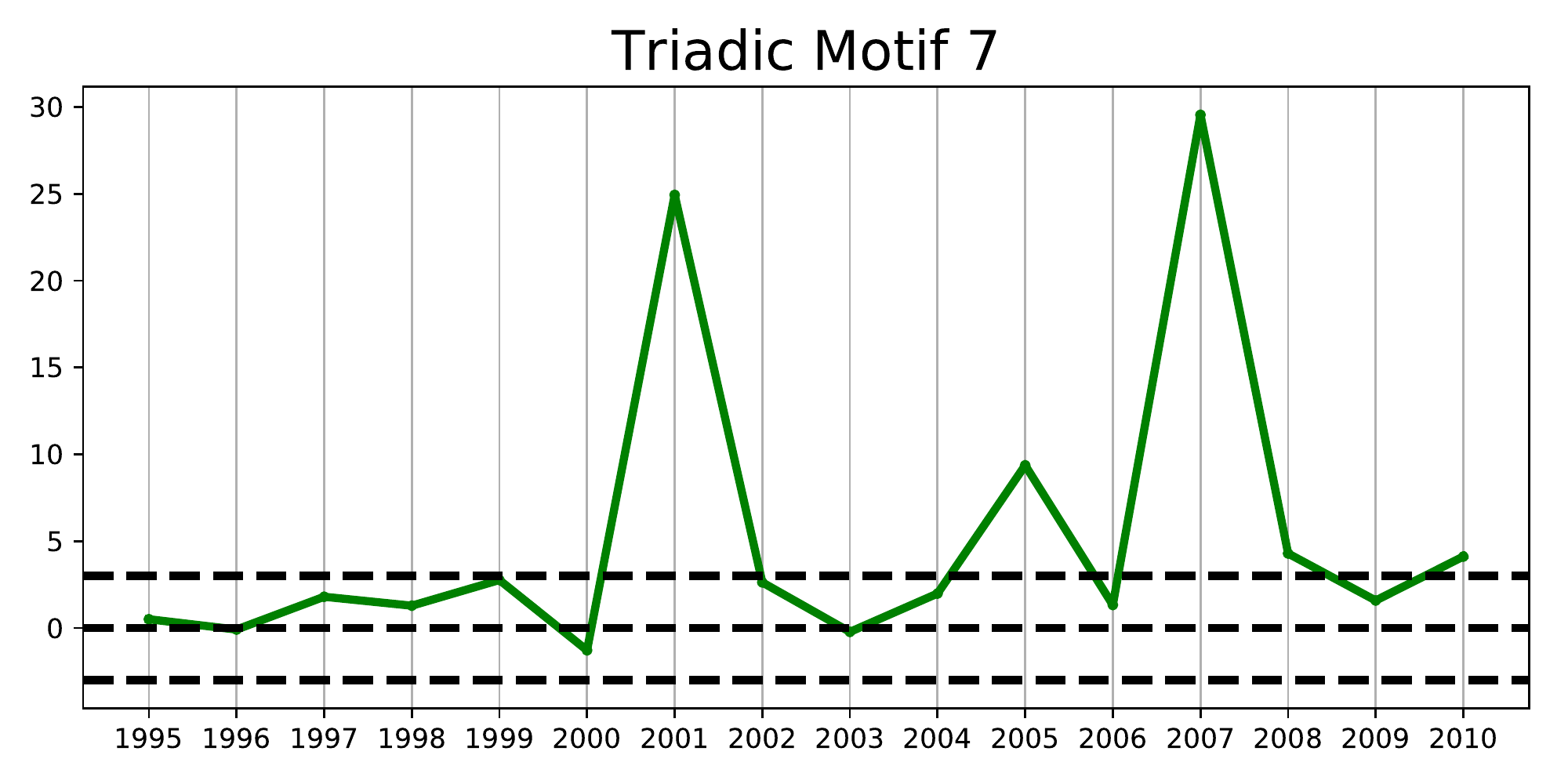}}
	}
	\makebox[\textwidth][c]{
    \subfloat[]
    {\includegraphics[scale=0.3]{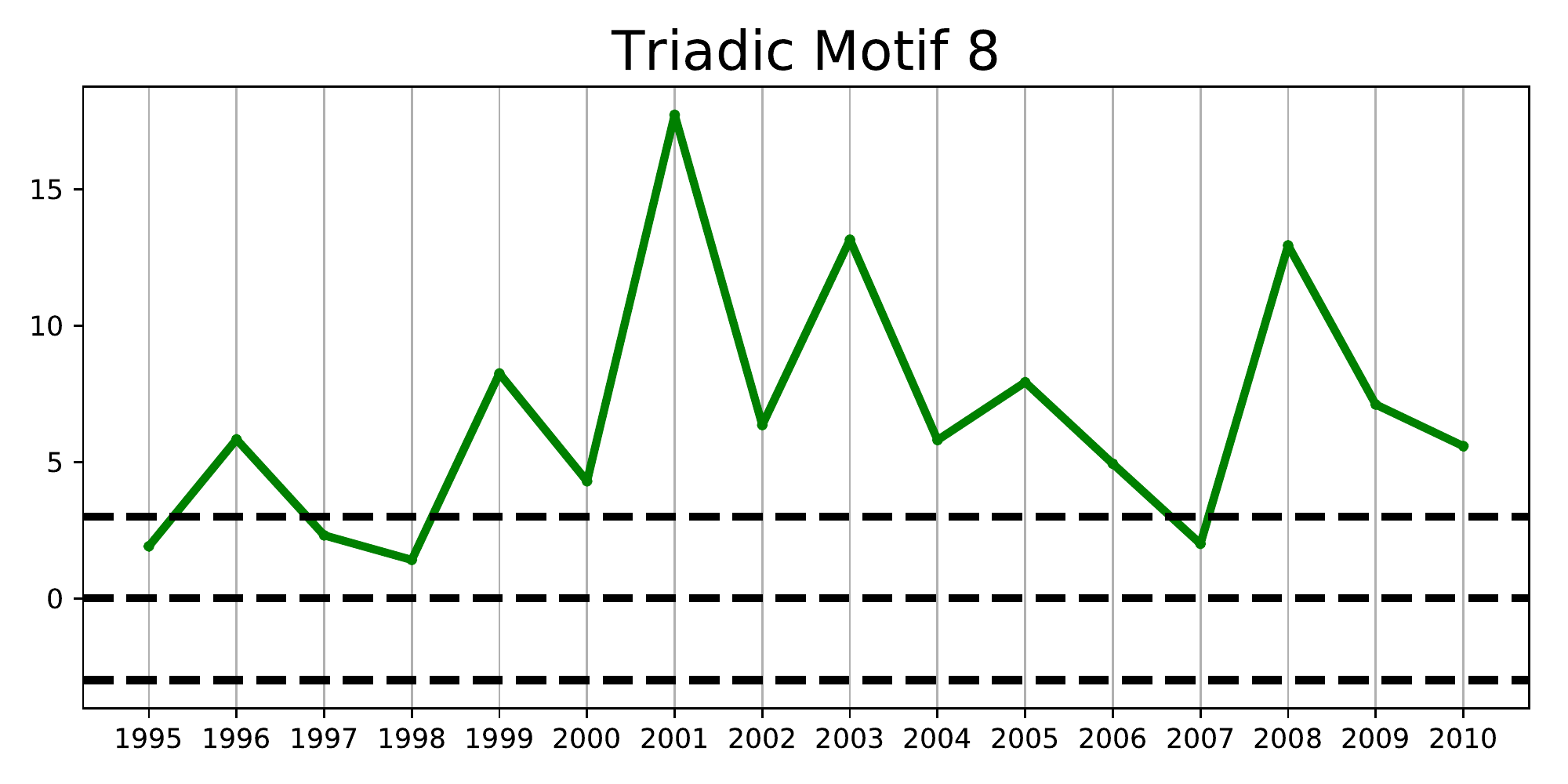}}
    \subfloat[]
	{\includegraphics[scale=0.3]{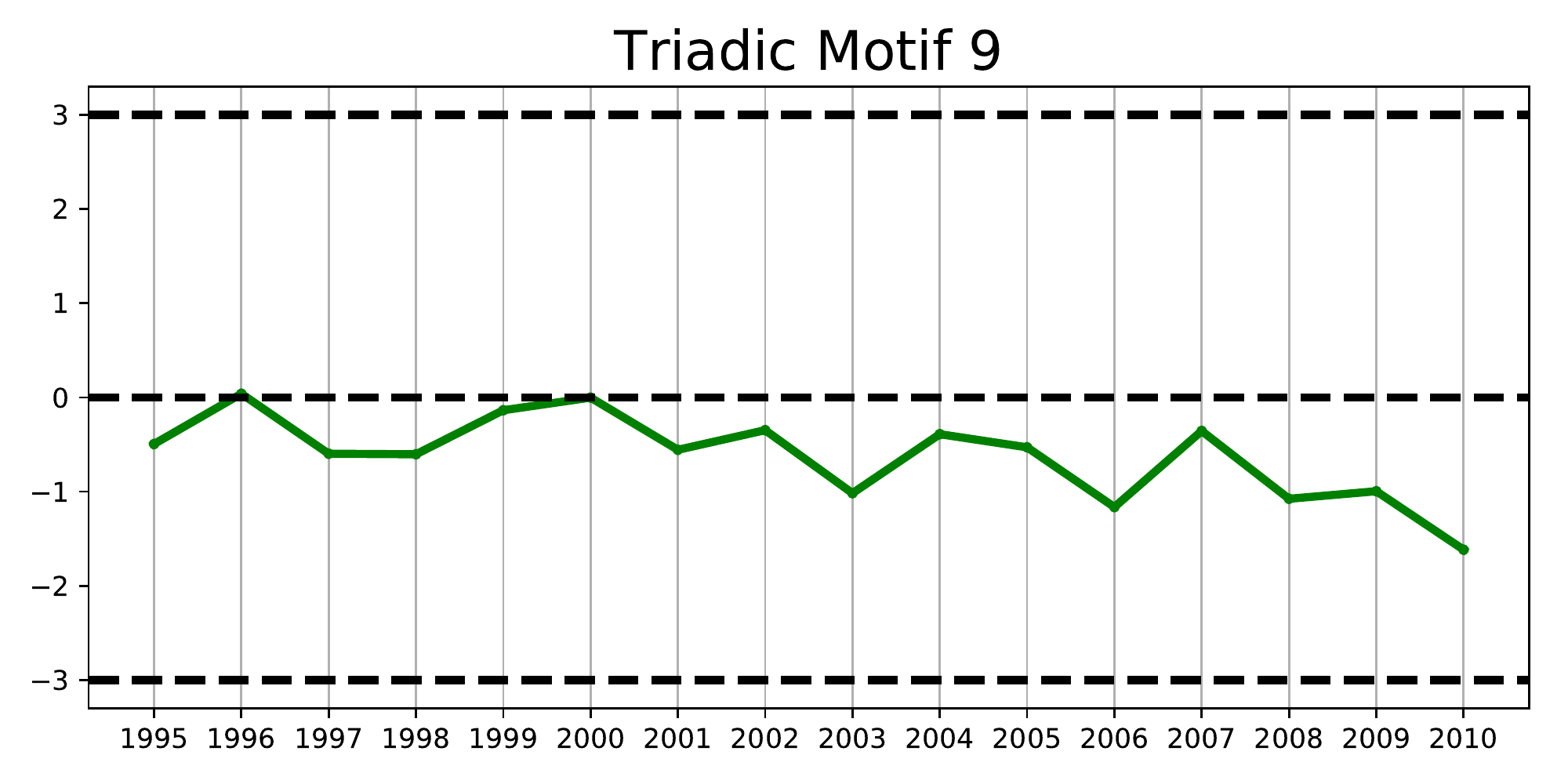}}
	}
	\makebox[\textwidth][c]{
    \subfloat[]
    {\includegraphics[scale=0.3]{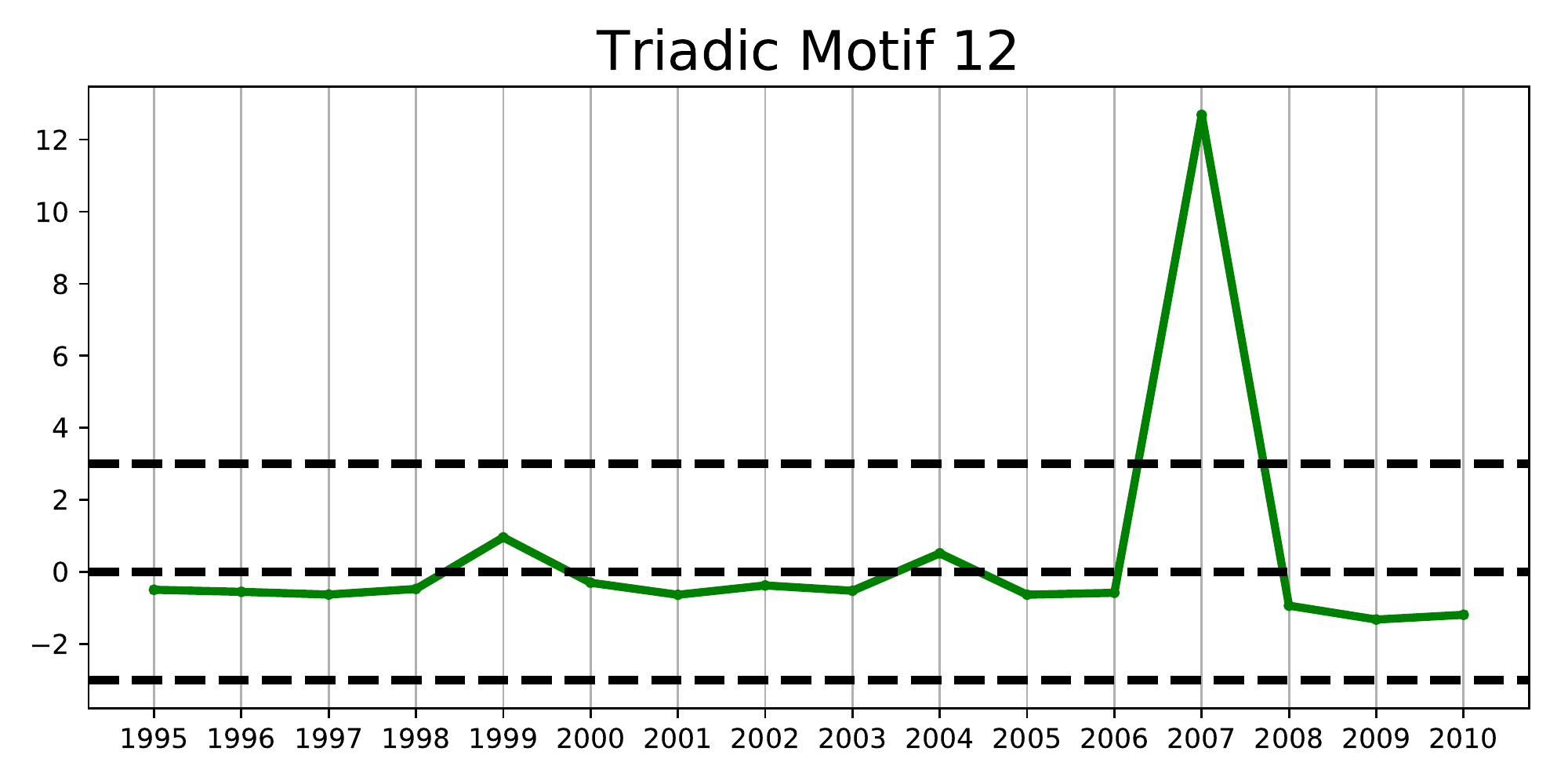}}
    \subfloat[]
	{\includegraphics[scale=0.3]{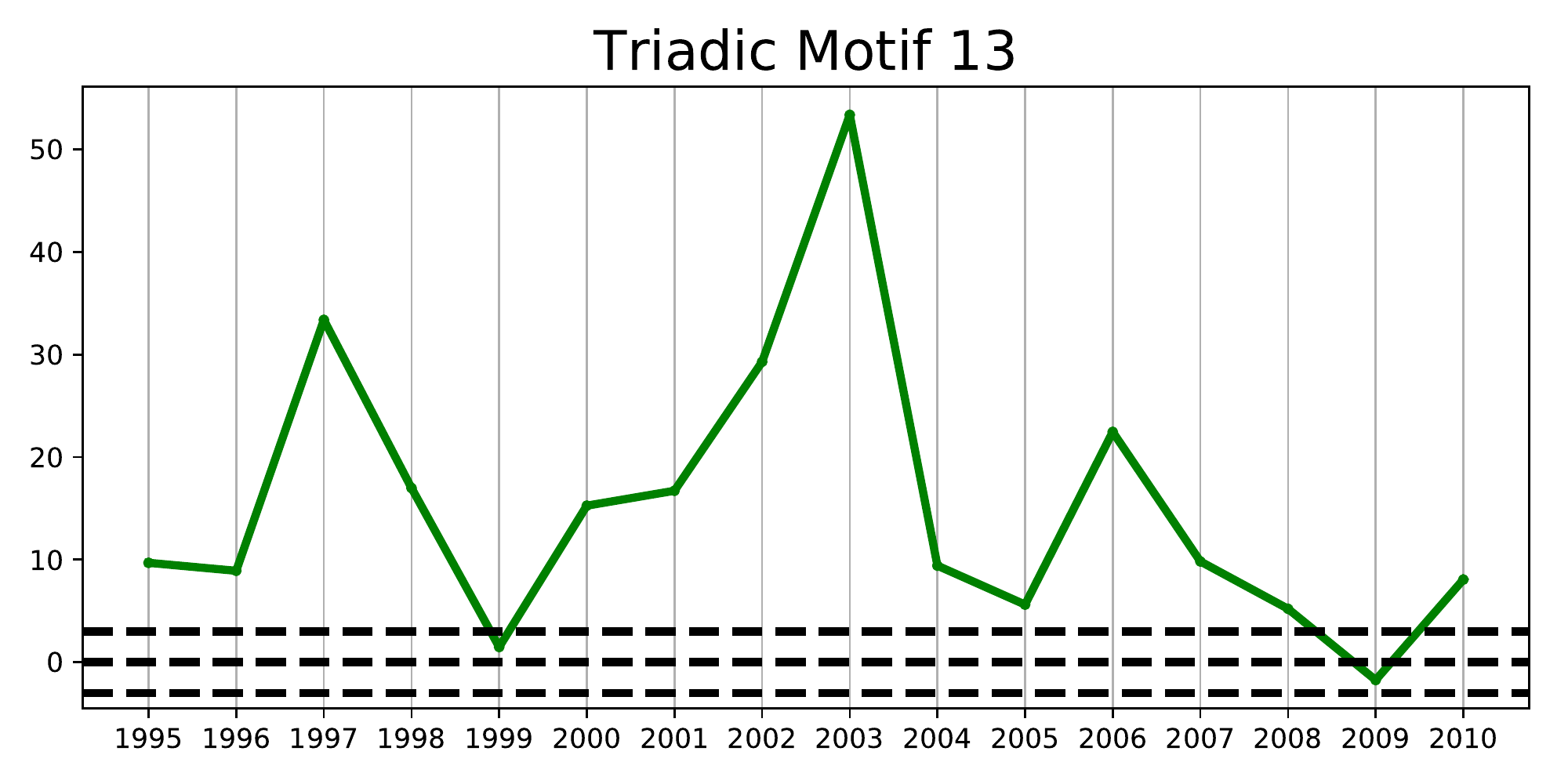}}
	}
    \end{center}
    \caption{Temporal series of z-score statistics for the triadic motifs in the weighted case.}
	\label{fig:triadsw}
\end{figure}

With the FiCM we can study how much the information about the volume of acquisitions affects the relevance of triadic motifs in the network. In Fig.~\ref{fig:triadsw} we can highlight that weighted motifs including reciprocated links are all underestimated by the ensemble model. Some motifs are highly underestimated before or after (3-12) the volume peaks, motif 8 presents a significant pattern during the entire time period between the two crisis, and  motifs 7-13 are underestimated in correspondence of the peaks or between the two.

\captionsetup[subfigure]{labelformat=empty}

\captionsetup[subfigure]{labelformat=empty}


\captionsetup[subfigure]{labelformat=empty}

\section{Conclusions}

In this paper we have explored the higher order pattern organization of the worldiwide web of Mergers\&Acquistions looking at the triadic motifs formation both at binary at weighted level.
The M\&A appears strongly asymmetrical with a low density, nevertheless it exhibits a high level of reciprocity in the whole period. Furthermore, the total volume of investments is chracterized by a wave-like behaviour, with two weaks in the period under study: one in 2000 and the second in 2007. 

The binary triadic motifs analysis reveals that reciprocity is highly relevant for this economic network. Indeed, the configuration model systemically underestimates the presence of reciprocal links (dyad) with the effect of underestimate also all motifs including only reciprocal links (motif n. 8, 12 and 13). On the other hand, motifs without reciprocal links are overestimated especially after the first wave of 2000. From the economic point of view, this could be interpreted as a specific feature of the M\&A web, meaning that there is a high  probability that country $i$ sending investments to country $j$ is receiving investments from country $j$. This could be interpreted as a preference of the country to send investments to a country from which it already receives them, for trust motivations or simply because they have already set specific agreements for the existing investment. Furthermoer, this says that the second wave of volume appears characterized by a decrease in the existence of one-direction link in favour of reciprocated ones. This outcome is not present in the first wave and this could shed light on the different nature of the two peaks of M\&As in the period under study. Since the reciprocity is crucial for this network, as we could expect fixing it as constraints, almost all triadic motifs are well reproduced. The only exception is represented by the motif number 8, overestimated by the RCM. This suggests that maybe fix the number of reciprocated link at node level is an hypothesis too strong that can be weakened with a novel constraint taking into account more than simple the number of node neighbours.

Moving to the weighted case, the scenario changes with motifs n. 7 and 8 highly underestimated from the null-model just close to the 2 volume waves, while n. 12 appears underestimated only around the second wave. On the other hand, motif 13 is underestimated between the two waves of investments.  
These outcomes suggest that weights shapes in a not trivial way the higher-order patterns observed in the M\&A web, revealing new features with respect to the binary part. For this reason, we think further work will be useful in this direction, trying to understand the economic reasons behind the emergence of some motifs just close to one or both investments waves, and exploring in more details the effect of reciprocity also in the weighted version of the network.

\section*{Acknowledgements}
This work is the output of the second edition of Complexity72h workshop, held at IMT School in Lucca, 17-21 June 2019 \url{https://complexity72h.weebly.com/}. Jian-Hong Lin also acknowledges the support from the China Scholarship Council (no. 2017083010177).

\bibliographystyle{ieeetr}
\bibliography{biblio}

\end{document}